\documentclass[natbib209,12pt,preprint]{aastex}

\usepackage{rotating}
\usepackage{graphicx}
\usepackage{epstopdf}

\newcommand{\msun}{$M_\odot$}
\newcommand{\rsun}{$R_\odot$}
\newcommand{\lsun}{$L_\odot$}
\newcommand{\kms}{km\,s$^{-1}$}

\newcommand{\ace}{$\alpha$}

\begin{document}

\title{Simulating the Common Envelope Phase of a Red Giant Using SPH and Uniform Grid Codes}
\author{Jean-Claude Passy$^{1,2}$, Orsola De Marco$^3$, Chris L. Fryer$^4$, Falk Herwig$^2$, Steven Diehl$^4$, Jeffrey S. Oishi$^5$, Mordecai-Mark Mac Low$^1$, Greg L. Bryan$^6$, Gabriel Rockefeller$^4$}

\altaffiltext{1}{Department of Astrophysics, American Museum of Natural History, New York, NY, USA}
\altaffiltext{2}{Department of Physics and Astronomy, University of Victoria, Victoria, BC, Canada}
\altaffiltext{3}{Department of Physics and Astronomy, Macquarie University, Sydney, NSW, Australia}
\altaffiltext{4}{Computational Computer Science Division, Los Alamos National Laboratory, Los Alamos, NM, USA}
\altaffiltext{5}{Kavli Institute for Particle Astrophysics and Cosmology, Stanford University, Palo Alto, CA, USA}
\altaffiltext{6}{Department of Astronomy, Columbia University, New York, NY, USA}

\begin{abstract}
We use three-dimensional hydrodynamical simulations to study the rapid infall phase of the common envelope interaction of a red giant branch star of mass equal to 0.88~\msun \ and a companion star of mass ranging from 0.9 down to 0.1~\msun. We first compare the results obtained using two different numerical techniques with different resolutions, and find overall very good agreement. We then compare the outcomes of those simulations with observed systems thought to have gone through a common envelope. The simulations fail to reproduce those systems in the sense that most of the envelope of the donor remains bound at the end of the simulations and the final orbital separations between the donor's remnant and the companion, ranging from 26.8 down to 5.9~\rsun, are larger than the ones observed. We suggest that this discrepancy vouches for recombination playing an essential role in the ejection of the envelope and/or significant shrinkage of the orbit happening in the subsequent phase.

\end{abstract}

		 


\section{Introduction}
\label{sec:intro}

Around 60\% of F and G stars are binaries, of which about 30\% have separations smaller than 30 AU and will interact during the primary's evolution \citep{Duquennoy1991}. During the giant phases of the primary, companions closer than $\sim 5$ AU enter a strong interaction phase with the primary and, under certain circumstances, a common envelope (CE) may form around the two stars. The secondary star spirals inside the envelope of the primary and may also fill its own Roche lobe because it cannot accrete all the matter coming from the donor star. This process is called a \textit{common envelope interaction} and was originally described by \cite{Paczynski1976}. For a general review of the topic see, e.g., \cite{Iben1993}. There are two different processes leading to the onset of a CE phase: the start of unstable mass transfer from the expanding primary to the secondary \citep{HjellmingWebbink1987, HurleyEtAl2002} and the development of a tidal instability that occurs if there is not enough angular momentum in the orbit to maintain the primary's envelope in synchronization \citep{Darwin1879}. The post-CE system will be either a compact binary system, if there is enough energy to eject the primary's envelope, or a merger, if not.

The CE interaction is an essential ingredient for any binary population synthesis study of intermediate \citep[e.g.,][]{PolitanoEtAl2010} or massive stars \citep[e.g.,][]{Belczynski2008}. Compact binaries are believed to be formed through at least one CE phase. Among them are symbiotic binaries, supersoft X-ray sources, cataclysmic variables and double white dwarfs, which are all possible supernova Type Ia progenitors. As \cite{MengEtAl2010} pointed out, results deduced from population synthesis studies such as the Type Ia supernova birth rate are highly dependent on the physics of the CE phase. Therefore, it is paramount to understand more accurately the CE interaction in order to identify the formation channels of such supernovae and to compare observations with predictive models. Moreover, many substellar companions to evolved stars have recently been discovered with small orbital separation. \cite{Maxted2006} found a brown dwarf orbiting a white dwarf with a 116 min period, while \cite{SetiawanEtAl2010} discovered a system composed of a Jupiter-like object orbiting an horizontal branch star with a 16.2 days period. We therefore know that substellar companions can survive a CE interaction, but what is the minimum mass of the companion that can eject the envelope? Is the ejected envelope entirely unbound or will some of it eventually fall back and form a circumbinary disk? Were the substellar companions present before and survived the CE or were they formed later on in such a disk \citep{Perets2010}? Those questions remain unanswered. 

Although the CE process was outlined more than 30 years ago, it is still far from understood quantitatively. Numerical simulations suggest that the typical duration of the entire CE phase is short --- less than $10^3$ years --- which makes CE ejections unlikely to be observed. However, one can use observations of post-CE binaries to better understand CE evolution. With the use of stellar models, the initial configuration of such systems can be approximately determined from the final configuration. Using either the \ace-formalism \citep[][but see \citealt{AlphaPaper2011}]{Webbink1984} or the $\gamma$-formalism \citep{Nelemans2000} the relevant parameters can be constrained and the CE ejection efficiency can be predicted. Using this approach, \cite{AlphaPaper2011} suggested an anti-correlation between $\alpha$, the CE efficiency parameter, and the secondary to primary mass ratio.

The entire CE evolution can be divided into three different phases \citep{Podsiadlowski2001} with different timescales, length scales and physics involved. These differences are the reasons why reproducing the entire CE evolution of a given system accurately is challenging. Therefore, one usually treats one phase after the other with different methods. In this paper, we focus only on the rapid infall phase, which has a short timescale ($\sim 1-10$ years), and in which the evolution is driven by drag forces. Several numerical hydrodynamic studies of the CE interaction have been carried out in the past \citep[for an exhaustive list, see][]{TaamSandquist2000}, including a series of ten papers starting with the two-dimensional calculation of the interaction of a 16~\msun \ supergiant and a 1~\msun \ neutron star \citep{PaperII}, and most recently treating three-dimensional simulations of the CE interaction between 3 or 5~\msun \ giant stars and 0.4 or 0.6~\msun \ main sequence (MS) companions \citep{Sandquist1998}. The latter study has been extended first by \cite{SandquistEtAl2000} to 1~\msun \ and 2~\msun \ red giant branch (RGB) stars with companion masses ranging from 0.1 to 0.45~\msun, then by \cite{DeMarco2003} to a 1~\msun \ asymptotic giant branch star with a 0.1 or 0.2~\msun \ companion. \cite{RickerTaam2008} computed high resolution simulations of the CE phase between a 1.05~\msun \ RGB star and a 0.6~\msun \ compact companion, and concluded that the gravitational component of the drag dominates over the hydrodynamical component \citep[also see][]{TaamRicker2010, RickerTaam2011}.

A direct comparison of the results obtained using different numerical methods has however never been carried out. Although analytical/empirical work has included discussion regarding observational data, there are only a couple of publications that connect simulations and observations in a meaningful way \citep[see e.g.,][]{SandquistEtAl2000}. Those are, as we will explain in \S\ref{sec:code} and \S\ref{sec:results}, key steps to better understand the implications of CE interactions and the physical processes driving them. In this paper we therefore present numerical simulations with two different algorithms of the CE interaction of a 0.88~\msun \ RGB star with a MS companion. Different companion masses from 0.1~\msun \ to 0.9~\msun \ are considered. The simulations are carried out with both an Eulerian code ({\it Enzo} in uniform-grid mode,  \cite{OsheaEtAl2004} and {\it enzo.googlecode.com}) and a Lagrangian code ({\it SNSPH}, \citealt{FryerEtAl2006}), and for different resolutions. We describe the numerical methods and the initial conditions of our 15 simulations in \S\ref{sec:code} and \S\ref{sec:simus}. We describe and discuss the results in \S\ref{sec:results} and \S\ref{sec:discussion}, and finally conclude and summarize in \S\ref{sec:conclu}.


\section{Description of codes}
\label{sec:code}

In this section we describe the numerical methods we use. We first compare the code algorithms and explain why a code-to-code comparison is necessary. Then, we describe both codes in detail and finally discuss different ways to compare resolution.

\subsection{Eulerian vs Lagrangian codes}
\label{subsec:codesdiff}

Although they are meant to simulate similar astrophysical situations, high order Eulerian grid codes and Lagrangian smoothed-particle hydrodynamics (SPH) codes differ fundamentally, with each having advantages and disadvantages. Among other studies, \cite{DaviesEtAl1993}, \cite{SantaBarbara}, \cite{AgertzEtAl2007}, \cite{TaskerEtAl2008} and \cite{HeitschEtAl2011} aim at identifying these differences. On the one hand, high-order Eulerian grid codes have a better wavenumber resolution than SPH codes for an equal number of cells and particles and are more accurate at resolving the rarefied regions since, unlike SPH, the resolution does not depend on the density of the gas; Eulerian codes also better resolve shocks \citep{TaskerEtAl2008} compared to SPH codes; and finally, SPH noise dominates subsonic flows and therefore makes it difficult for SPH codes to follow perturbations in flows with Mach numbers under unity. On the other hand, SPH codes don't diffuse material properties, and inherently conserve mass, momentum and energy \citep{Rosswog2009}. While the treatment of boundary conditions can be challenging in grid-based codes when the flow expands beyond the computational domain, SPH easily handles vacuum conditions. It is still unclear which method is the most appropriate to simulate CE interactions. Therefore, we use both methods and confront the results from both codes in order to draw conclusions about their physical relevance.

\subsection{Input physics}
\label{subsec:input}

Both codes solve the fully compressible hydrodynamics equations with self-gravity included. These equations can be written using an Eulerian formulation:

\begin{equation}
	\frac{\partial \rho}{\partial t} + \nabla \cdot (\rho \mathbf{v}) = 0	
	\label{eq:hd1}
\end{equation}
  
\begin{equation}
	\frac{\partial \mathbf{v}}{\partial t} + (\mathbf{v} \cdot \nabla) \mathbf{v}  = - \frac{1}{\rho} \nabla p - \nabla \Phi  
	\label{eq:hd2}
\end{equation}

\begin{equation}
	\frac{\partial u}{\partial t} + \mathbf{v} \cdot \nabla u = - \frac{1}{\rho} \nabla \cdot (p\mathbf{v}) - \mathbf{v} \cdot \nabla \Phi
	\label{eq:hd3}
\end{equation}

\begin{equation}
	u_{int} = \frac{1}{\gamma - 1} \frac{p}{\rho} 
	\label{eq:hd4}
\end{equation}

\begin{equation}
	\Delta \Phi = 4 \pi G \rho 
	\label{eq:poisson}
\end{equation}
 
\noindent where $\rho, \mathbf{v}, p, \Phi, u, u_{int}, \gamma $ are the density, velocity, pressure, gravitational potential, specific total energy, specific internal energy and adiabatic index of the gas, respectively. The total energy is the sum of the internal energy, and the macroscopic kinetic energy:

\begin{equation}
	u = u_{int} + \mathbf{v}^2 / 2 .
	\label{eq:energy}
\end{equation}

\noindent Equations (\ref{eq:hd1}), (\ref{eq:hd2}) and (\ref{eq:hd3}) express mass continuity, conservation of momentum and conservation of energy, respectively. Both codes evolve the internal energy rather than the total energy. An ideal gas equation of state (Eq.~\ref{eq:hd4}) for a monoatomic gas ($\gamma = 5/3$) closes the system composed by equations (\ref{eq:hd1})-(\ref{eq:hd3}). Such an equation of state represents an adequate approximation of the deep convective envelope of RGB stars \citep{HjellmingWebbink1987} although it ignores some physical processes such as radiation pressure and ionization. We discuss this point in detail in \S\ref{subsub:missing}. Finally, the gravitational potential is calculated using the Poisson equation (Eq.~\ref{eq:poisson}).

\subsection{The {\it Enzo} code}
\label{sub:Enzo}

{\it Enzo} is a three-dimensional, adaptive mesh refinement hybrid (hydrodynamics + N-body) grid-based code \citep[]{BryanEtAl1995, OsheaEtAl2004} that we use in uniform-grid mode only. It is primarily designed to simulate cosmological structure formation \citep{NormanEtAl2007}. However, its numerous features make it useful for reproducing many different astrophysical situations, including CE interactions.

The Euler equations (Eqs.~\ref{eq:hd1}--\ref{eq:hd3}) are solved using the \cite{VanLeer1977} second-order advection method also implemented in {\it Zeus} \citep{StoneNorman1992}. Although those equations can also be solved in {\it Enzo} by a third-order piecewise parabolic method that better resolves shocks and turbulence, our tests show that it slows down the computation by a factor 2. As we will point out in \S\ref{sec:results}, there are neither strong shocks nor important turbulence in our simulations so we favor efficiency and use the van Leer solver. The Poisson equation is solved using fast-Fourier transforms.

In the case of a CE interaction between a RGB star and a MS companion, the radius of the secondary --- typically 0.5 \rsun \ --- is small compared to the primary's radius ($\sim 100 $ \rsun), so we can legitimately model the companion as a point mass particle. Furthermore, as shown in Fig.~\ref{fig:profiles}, the primary's core is also small ($\sim 0.01 $ \rsun) and dense, so it can also be modeled as a point mass. 

{\it Enzo} usually models collisionless particles as a continuous mass field appropriate for computing the gravitational potential in the case that each particle represents many actual particles, such as in cosmological simulations with dark matter. In that case their mass is deposited in the 8 nearest cells and added to the gas density of those cells to find the total density for use in solving the Poisson equation (Eq.~\ref{eq:poisson}). In a simple two-body interaction between 1~\msun \ and 0.1~\msun  \ objects in a one year circular orbit without gas, this method does not provide the accuracy required by our problem because of the spreading out of the mass of the point source, leading to an inaccurate gravitational potential. Indeed, a 1 \% error in the orbit is reached after only 6 orbits. Consequently, we implemented, as a new type of particle, point mass particles. These particles create a potential that is added analytically to the gas potential calculated using the Poisson equation. Using an analytic potential yields an accuracy of the orbit more than two orders of magnitude better than with the default particles. The gravitational potential created by a point mass particle is smoothed according to the prescription of \cite{Ruffert1993}, used in \cite{Sandquist1998}:

\begin{equation}
	\Phi_{PM}(r) = \frac{-GM_{PM}} { \sqrt{r^2 + \epsilon^2 \delta^2 \exp{[-r^2/(\epsilon\delta)^2]}}}
\end{equation}

\noindent where $M_{PM}$ is the mass of the particle, $r$ is the distance from the particle, $\delta$ is the size of a cell and $\epsilon = 1.5$. The point mass particles are advanced using a leapfrog algorithm. Time stepping is determined by taking the minimum time step between the Courant conditions for the gas, the particles and the acceleration field:

\begin{equation}
	\delta t_{gas} =  \min_{\rm cells}\left(\frac{C_1 \delta}{c_s + \max(|v_x|,|v_y|,|v_z|)}\right)
	\label{eq:cond1}
\end{equation}

\begin{equation}
	\delta t_{part} = \min_{\rm particles}\left(\frac{C_2 \delta}{\max(|V_x|,|V_y|,|V_z|)}\right)
	\label{eq:cond2}
\end{equation}

\begin{equation}
	\delta t_{accel} =  \min_{\rm cells}\left( \sqrt{\frac{\delta}{\max(|g_x|,|g_y|,|g_z|)}}\right)
	\label{eq:cond3}
\end{equation}

\noindent where $C_1 = 0.4$ is the Courant factor, $C_2 = 0.4$ is the particle Courant factor, $c_s$ is the sound speed, $\mathbf{v} = (v_x,v_y,v_z)$ is the velocity of the gas, $\mathbf{V} = (V_x,V_y,V_z)$ is the velocity of a particle and $\mathbf{g} = (g_x,g_y,g_z)$ is the acceleration field.

Finally, we remark that the current {\it Enzo} Poisson solver prevented us from using nested or adaptive grids that would have allowed us to increase resolution locally. The inaccurate treatment of boundary conditions within the refined grids prevented us from stabilizing the RGB progenitor in a multi-grid initial setup. We are currently developing a new Poisson solver that will allow us to use nested grids as well as adaptive mesh refinement and carry out better-resolved simulations.

\subsection{The {\it SNSPH} code}
\label{sub:sph}

\begin{figure}[h!]
	\begin{center}
		\includegraphics[scale=0.4]{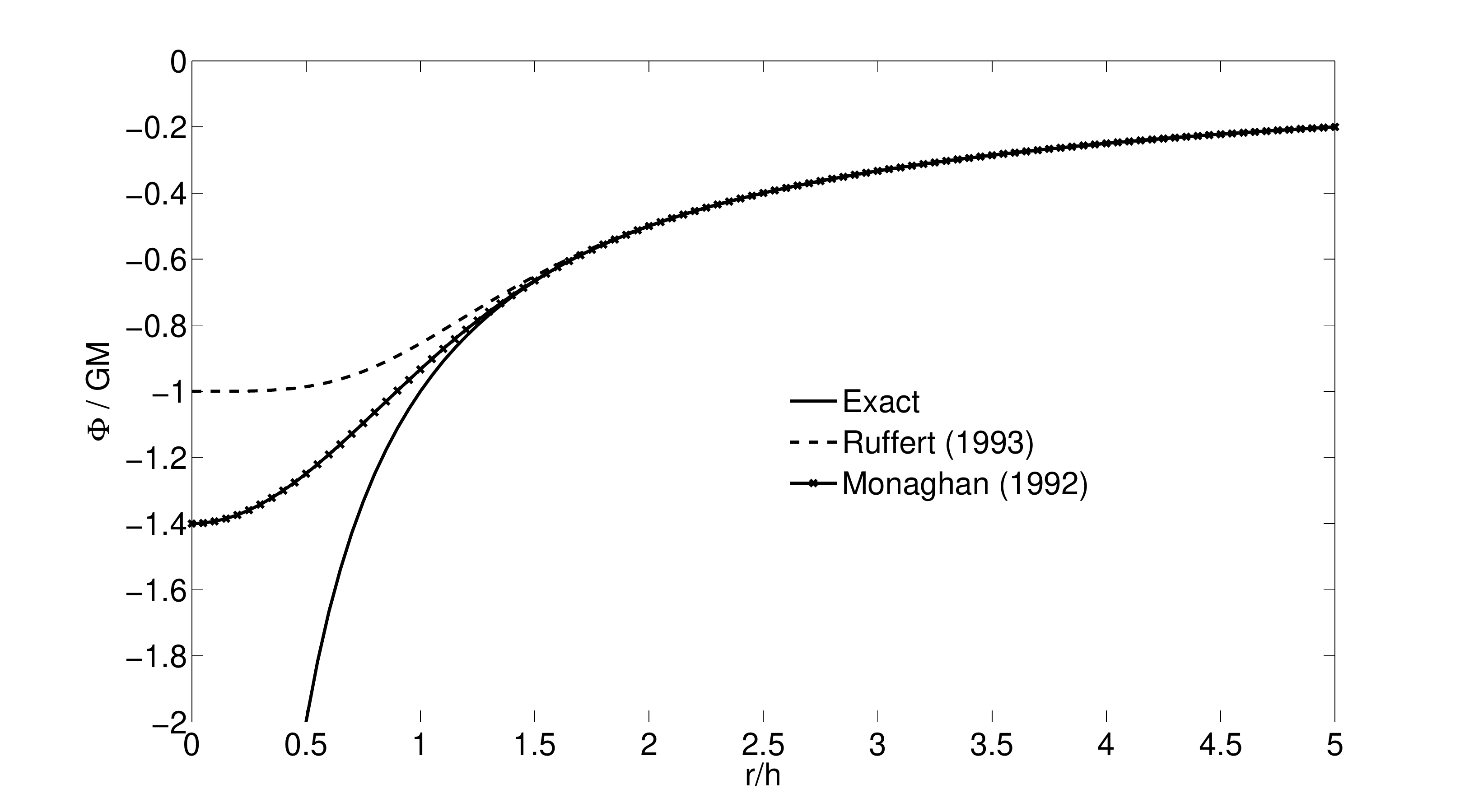}		
	\caption{Comparison between the different potentials in arbitrary units with $h_i=\epsilon \delta = 1$. Plotted are the theoretical potential (solid line), the \cite{Ruffert1993} potential used in {\it Enzo} (dashed line) and the \cite{Monaghan1992} one used in {\it SNSPH} (dash-cross line).
	\label{fig:potential}
	}
	\end{center}
\end{figure}

{\it SNSPH} \citep*{FryerEtAl2006} is a three-dimensional, parallel SPH code using tree gravity. It uses a regular Monaghan cubic spline kernel \citep{Monaghan1992}. For the artificial viscosity we use the sum of a bulk viscosity and a von Neumann and Richtmyer viscosity \citep{Rosswog2009}. The particles are organized into a parallel hashed oct-tree as described in \cite{WarrenSalmon1993}.
The gravitational potential of a SPH particle, $i$, is smoothed using the following formula:

\begin{equation}
	\Phi_i (x_i = r_i/h_i) = \left\{
	\begin{array}{ll}
	-G m_i / h_i  \times (\frac{2}{3} x_i^3 - \frac{3}{10} x_i^4 +  \frac{1}{10} x_i^5 - 1.4)   \hspace{2.9cm} \rm \ \  if \ \ \ 0 \leq x \leq 1 \\ \\
	-G m_i / r_i  \times \left[(\frac{4}{3} x_i^2  - x_i^3 + \frac{3}{10}x_i^4 - \frac{1}{30} x_i^5 - 1.6)/h_i + 1/15r_i \right] \ \rm \ \  \  \ \ if \ \ \ 1 \leq x \leq 2 \\ \\ 
	-G m_i / r_i \ \ \ \rm  \hspace{6.5cm} otherwise\\
	\end{array}
	\right.
\end{equation}

\noindent where $h_i$, $m_i$ and $r_i$ are the smoothing length, the particle mass and the distance from the particle, respectively. We compare both numerical potentials to the theoretical potential in Fig.~\ref{fig:potential}. For a given smoothing length, $h_i$, the \cite{Monaghan1992} potential used in our {\it SNSPH} simulations is deeper than the \cite{Ruffert1993} one used in our {\it Enzo} simulations. Also, the \cite{Monaghan1992} potential is exact at distances larger than $2 h_i$ whereas the \cite{Ruffert1993} potential only asymptotically tends to the exact potential. 

SNSPH uses the fast multipole method to calculate gravitational accelerations \citep{WarrenSalmon1993}. The SPH particles are also advanced using an leapfrog algorithm. Finally, in order to keep the same overall spatial coverage, the smoothing length varies according to the formula from \cite{Benz1989}:

\begin{equation}
	\frac{h_i(t)}{h_i(0)} = \left(\frac{\rho_i(0)}{\rho_i(t)}\right)^{1/3} .
	\label{eq:benz1}
\end{equation}

\subsection{Resolution comparison}
\label{subsec:resol}

There is no ideal way to compare the resolution between SPH and uniform-grid codes. However, a few criteria can give us a general idea of how to relate them. 

As mentioned by \cite{DaviesEtAl1993}, a first global criterion would be to compare the total number of SPH particles $N_{part}$ with the total number of cells originally inside the progenitor:

\begin{equation}
	N_{cells} = \frac{V_1}{V_G} \times N_{tot}  \sim 4.19 \times \left(\frac{R_1}{L}\right)^3 N_{tot} , 
\end{equation}

\noindent where $N_{tot}$, $V_1$, $V_G$, $R_1$ and $L$ are the total number of cells, the volume of the primary, the volume of the grid, the radius of the primary and the linear dimension of the grid, respectively. As time goes by, the gas will however fill a larger fraction of the numerical grid and thus increase the number of relevant cells, but not the real resolution of the simulation. 

\begin{figure}[h!]
	\begin{center}
		\includegraphics[scale=0.14]{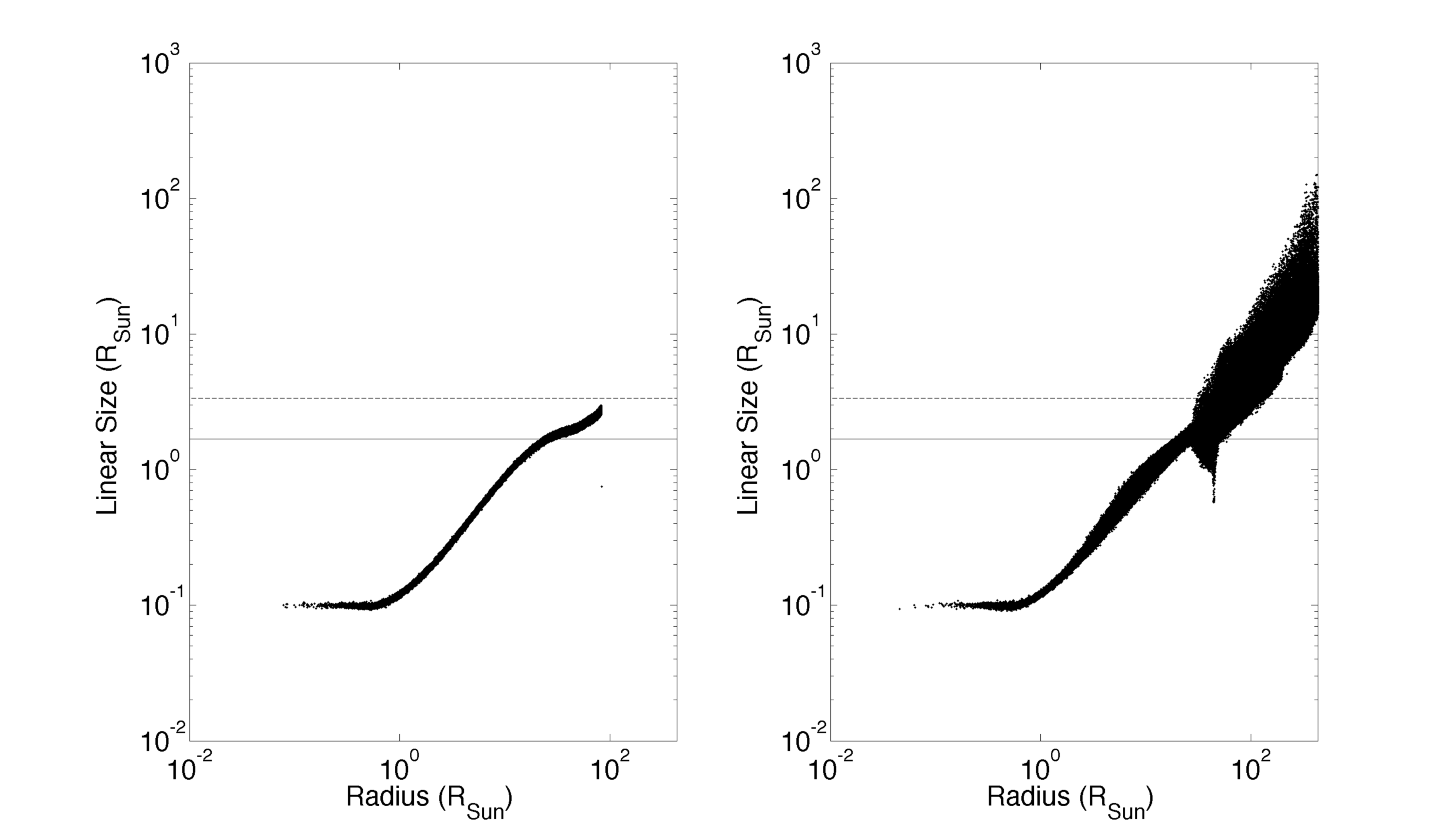}
	\caption{Resolution comparison between the {\it SNSPH} smoothing length field (dots) for a run with 500\,000 particles, and the {\it Enzo} size of a grid cell for the $128^3$ (dash line) and the  $256^3$ (solid line) runs for the initial (left) and final (right) particle distributions.
	\label{fig:h}
	}
	\end{center}
\end{figure}

A more local criterion is to compare the size of an {\it Enzo} grid cell, $\delta$, with the SPH smoothing length, which varies in space and time. Indeed, if the companion does not sink much into the primary's envelope and does not modify the inner part of the smoothing length distribution too much, then the resolution deep inside the progenitor does not matter. Therefore, we compare the smoothing length distribution of the SPH model to the cell size of the Eulerian grid. As shown in Fig.\ \ref{fig:h}, the smoothing length at small radii does not vary, so an {\it Enzo} run with a $128^3$ grid will be under-resolved compared to our canonical 500\,000 (roughly $80^3$) particle SPH run no matter how deep the companion penetrates while a run with a $256^3$ grid would be equivalent to our SPH runs if the separation between the primary core and the companion always exceeds 20 \rsun. This local criterion for the resolution is not perfect either since it does not take into account the variation of the smoothing length throughout the SPH simulation. 

Again, comparing the resolution between uniform-grid and SPH codes is quite challenging and both methods have, in the situation we are interested in, strengths and weaknesses: SPH will under-resolve the low-density outer parts of the envelope, where the smoothing length dramatically increases, while it will be more accurate in the later phase of the evolution when the separation between the primary's core and the secondary will typically sink below a few cells. Therefore, comparing SPH and grid-based simulations is paramount in order to state which one is more adapted to our problem, and the combination of both global and local criteria is the best way to compare the resolutions of both methods.

\section{The simulations}
\label{sec:simus}

We perform 5 {\it SNSPH} and 12 {\it Enzo} simulations of CE interactions with a 0.88 \msun \ RGB primary that are summarized in Table~\ref{tab:runs}. The {\it SNSPH} simulations are computed using $500\,000$ particles whose initial smoothing length follows the radial profile shown in Fig.~\ref{fig:h}. The {\it Enzo} simulations are performed using either a $128^3$ or a $256^3$ grid. In both cases, the linear size of the computational domain is $L = 3 \times 10^{13}$~cm. We consider companion masses of 0.9, 0.6, 0.3, 0.15 and 0.1~\msun. Giant stars are slow rotators with rotational velocities of the order of a few \kms \ \citep{DeMedeirosMayor1999}. Although it is expected that a close companion will, through the action of tides and the transport of angular momentum in the primary envelope, spin up the envelope during the pre-CE phase, the actual rotation of the primary at the onset of the CE interaction is hard to quantify. Moreover, even if the primary was uniformly rotating at 50 \kms, its rotational energy would be

\begin{equation}
	E_{rot} = \frac{1}{2} r_g M_1 R_1^2 \omega^2 \sim 2.2 \times 10^{44} \rm \ ergs
\end{equation}

\noindent where $\omega$, $r_g$, $M_1$ and $R_1$ are the angular velocity, the radius of gyration, the mass and the radius of the primary, respectively. For RGB stars $r_g$ is typically about 0.1 \citep{TaamSandquist2000}. This rotational energy does not affect the energetics of the system since it is more than two orders of magnitude smaller than the binding energy of the primary (see below). Consequently, we assume that the primary is initially non-rotating. Finally, the companion is at the start placed at the surface of the primary in a circular orbit. We thus have three different simulations for each initial companion mass - one with SNSPH, and two with {\it Enzo} on $128^3$ and $256^3$ grids. Additionally, we also run two {\it Enzo} simulations in order to study the dependency of the final parameters on the initial conditions. We consider the $128^3$ {\it Enzo} simulation with a 0.3~\msun \ companion (Enzo3) as the reference and run identical simulations increasing, by 5 \%, either the initial velocity of the companion (Enzo11) or the initial separation (Enzo12). All the runs follow the evolution of the system for about $1\,000$~days.

\begin{table}[h!]
\begin{center}
\scalebox{0.9}
	{\begin{tabular}{cccccccc}
	\hline
	\hline
	  & $N_{part} \ {\rm or} \ N_{cells}$ & $M_2$ (\msun) & $A_0$ (\rsun) & $P_0$ (days) & $v_0/v_{circ}$ & $A_f$ (\rsun) & $P_f$ (days) \\
	\hline
	 SPH1 & 500 000 &  0.9 &  83 & 66 & 1 & 26.8 & 13.5 \\
	 SPH2 & 500 000 &  0.6&  83 & 72 & 1 & 20.6 & 10.1 \\
	 SPH3 & 500 000 &  0.3&  83 & 81 & 1 & 11.3 & 5.5 \\
	 SPH4 & 500 000 &  0.15&  83 & 86 & 1 & 7.3 & 3.0 \\
	 SPH5 & 500 000 &  0.1&  83 & 88 & 1 & 6.1 & 2.2 \\
	\hline
	 Enzo1  & $128^3$ &  0.9 &  91 & 75 & 1 & 28.1 & 15.5 \\
	 Enzo2  & $128^3$ &  0.6 &  91 & 83 & 1 & 20.0 & 11.0 \\
	 Enzo3  & $128^3$ &  0.3 &  91 & 93 & 1 & 11.7 & 5.6 \\
	 Enzo4  & $128^3$ &  0.15 &  91 & 99 & 1 & 8.6 & 3.4 \\
	 Enzo5 & $128^3$ &  0.1 &  91 & 102 & 1 & 8.5 & 3.3 \\
	\hline
	 Enzo6  & $256^3$ &  0.9 &  85 & 68 & 1 & 25.5 & 13.2 \\
	 Enzo7  & $256^3$ &  0.6 &  85 & 75 & 1 & 19.2 & 9.8 \\
	 Enzo8  & $256^3$ &  0.3 &  85 & 84 & 1 & 11.2 & 5.4 \\
	 Enzo9  & $256^3$ &  0.15 &  85 & 89 & 1 & 6.9 & 2.8 \\
	 Enzo10  & $256^3$ &  0.1 &  85 & 92 & 1 & 5.7 & 2.1 \\
	\hline
	\hline
	 Enzo11  & $128^3$ &  0.3 &  91 & 93 & 1.05 & 12.0 & 4.6 \\
	 Enzo12  & $128^3$ &  0.3 &  95.5 & 99 & 1 & 12.2 & 5.0 \\
	\hline
	\hline
 	\end{tabular}}
	\caption{Main parameters for the different simulations. Reported are the number of particles ($N_{part}$) or cells ($N_{cells}$), the companion mass, the initial orbital separation ($A_0$), the initial orbital period ($P_0$), the ratio of the initial orbital velocity of the companion ($v_0$) to the velocity required for a circular orbit ($v_{circ}$) and the final orbital separation ($A_f$) taken at the end of the rapid infall phase (\S~\ref{subsec:infall}).
	\label{tab:runs}
	}
\end{center}
\end{table}

\begin{figure}[h!]
	\begin{center}
		\includegraphics[scale=0.083]{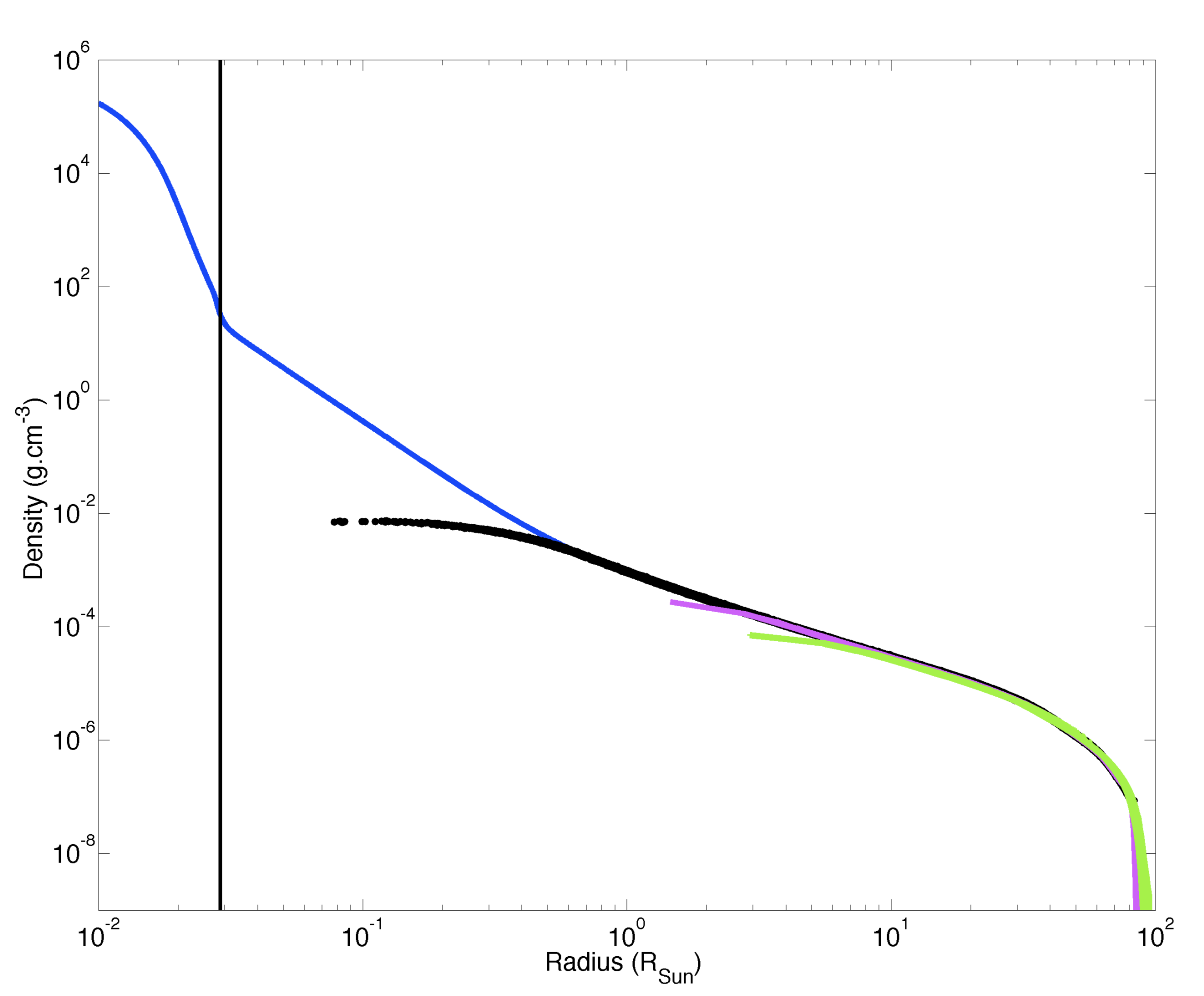}
		\includegraphics[scale=0.083]{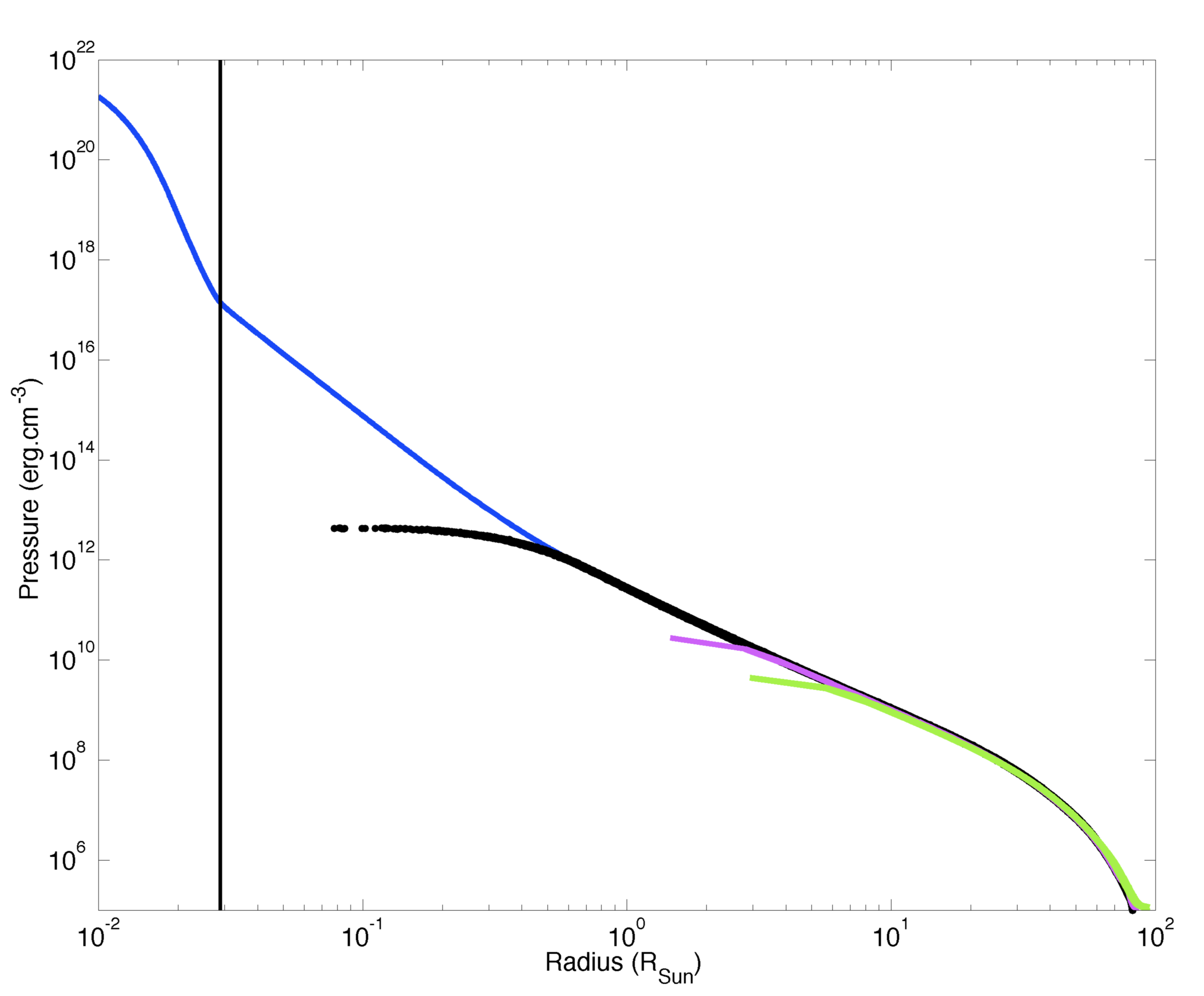}
		\includegraphics[scale=0.083]{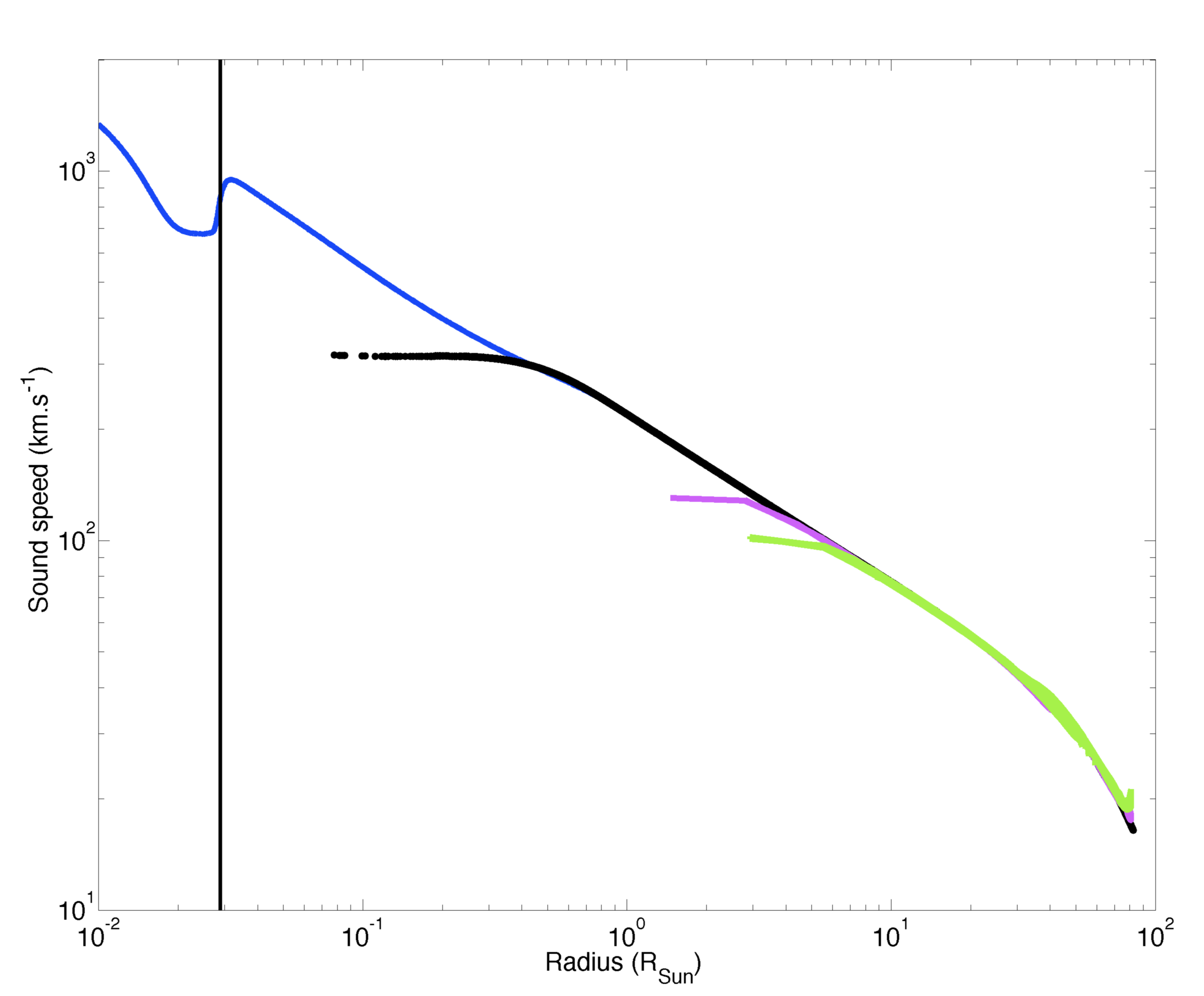}
		\includegraphics[scale=0.083]{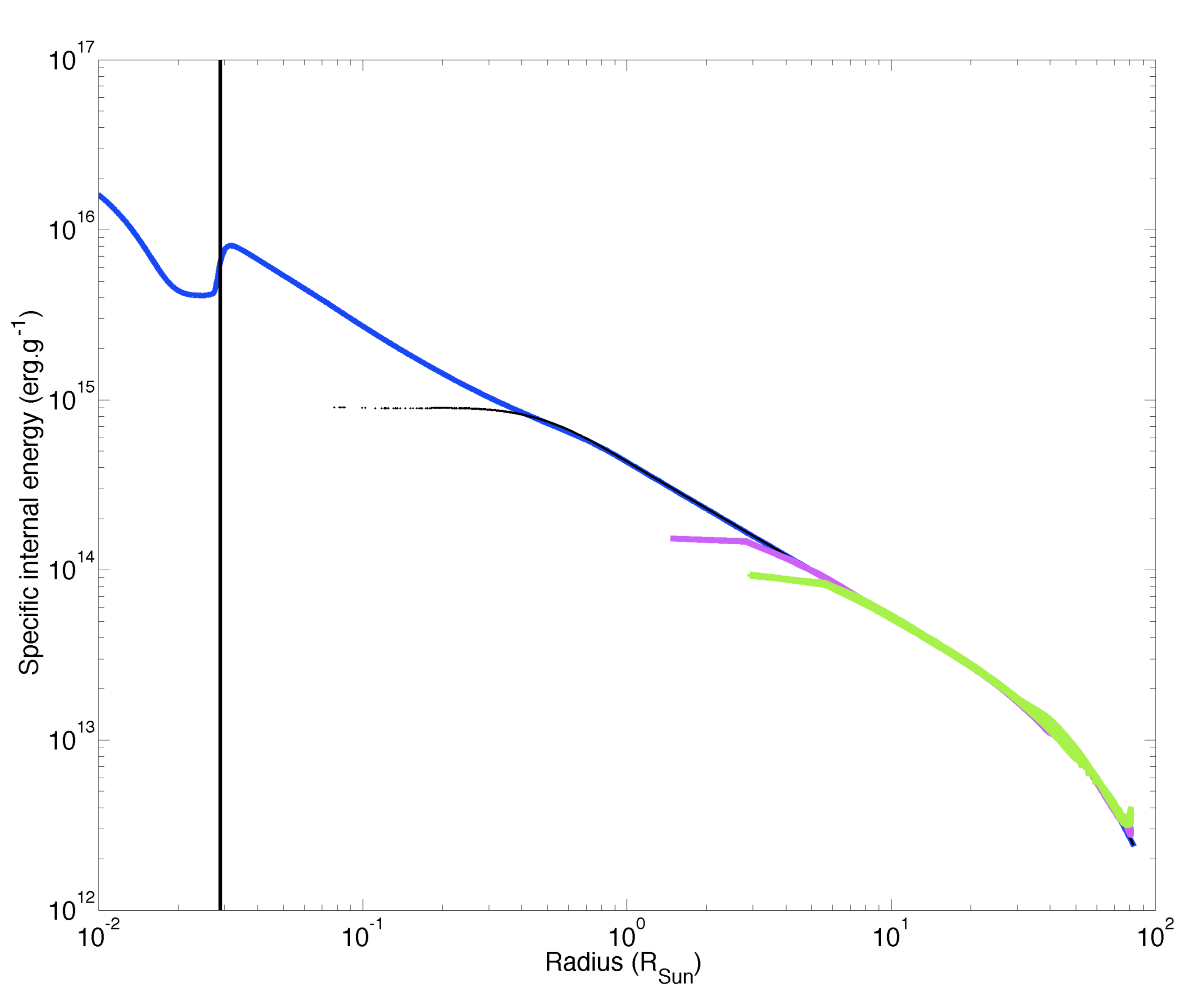}
	\caption{Comparison between the EVOL stellar evolution model (blue), the SPH initial model computed with 500 000 particles (black) and the {\it Enzo} initial models for a $128^3$ (green) and a $256^3$ (purple) unigrid. The vertical line represents the core-envelope boundary according to the criterion of \cite{AlphaPaper2011}. 	
	\label{fig:profiles}
	}
	\end{center}
\end{figure}

As a primary, we use a one-dimensional model of a star with a MS mass of 1~\msun. Using the stellar evolution code EVOL \citep{Herwig2000}, this progenitor was evolved to the RGB phase until the core reached $M_c = 0.392$~\msun. At that time, the radius of the star was 83~\rsun \ and its total mass was $ M_1 = 0.88$~\msun \ due to mass loss, which was treated using the Reimers formalism with $\eta = 0.5$. We adapt this model by using the density and pressure profiles, but computing the internal energy using Eq.~\ref{eq:hd4}. A sample of relevant profiles are plotted in Fig.~\ref{fig:profiles}. 

We now explain how this stellar model is modified in order to be compatible with an input suitable for each of our codes. For the {\it SNSPH} simulations, the initial particle configuration is a weighted Voronoi tesselation (WVT) similar to that described by \cite{DiehlStatler2006}. As we have explained in \S\ref{subsec:codesdiff} one limitation of SPH codes is the large number of particles required by dense regions such as the core of the primary. Since the time step induced by a particle $i$ can be roughly estimated by $h_i/c_{s,i}$ where $c_{s,i}$ is the local sound speed, a small smoothing length will require a small time step resulting in a high computational cost. Since the equation of state changes significantly around the helium core, we represent the core by a particle with mass $M_c$. The associated smoothing length is $h_c = 0.1$ \rsun. We add SPH particles in the region around the core such that the density values and gradient profiles connect smoothly at the core/envelope boundary ($ r = 2 h_c = 0.2$~\rsun). In this way, we obtain the profile shown in Fig.~\ref{fig:profiles}. Since the density profile has been changed, one must modify the gravitational acceleration accordingly. Assuming hydrostatic equilibrium in spherical symmetry, we integrate the pressure gradient choosing the integration constant to match the true profile outside the core (at $r = 0.2$~\rsun). The specific energy profile is computed using Eq.~\ref{eq:hd4}. Finally, the acceleration of a SPH particle is due either to gravity or to gas pressure. These two components are computed using the same particle mass for all particles except the core and the companion, for which we distinguish between the gravitational and SPH masses. The gravitational mass of the core is $M_c$ and its SPH mass is set to balance the gravitational acceleration of the envelope and prevent the star from collapsing. As for the companion, we treat it as an N-body particle so its SPH mass is 0 \msun.

For the {\it Enzo} simulations, the grid is initialized using the stellar model of the primary with the addition of a PM particle that represents its core. We fill the computational domain with a constant background density to prevent the star from expanding and set the ratio between the background density and the minimum density of a cell that belongs to the primary to $10^{-4}$. This setup is not initially numerically stable. The star tends to expand, so we let the initial configuration evolve for a few dynamical times in the absence of the companion, while damping the velocity field by a factor of 2 after each cycle. Finally, we evolve this relaxed model normally for another few dynamical times to obtain a numerically stable model. As a side effect of the relaxation to hydrostatic equilibrium, the {\it Enzo} models are a little bit bigger --- the lower the resolution, the larger the radius of the primary is --- thus the initial orbital separations between the models are slightly different (Table \ref{tab:runs}).

\section{Results}
\label{sec:results}

In this section, we describe the results obtained from our 15 simulations. Since the qualitative behavior is the same in all of them, we detail the 0.6~\msun \ case (SPH2, Enzo2 and Enzo7).

\subsection{Description of the rapid infall phase}
\label{subsec:infall}

\begin{figure}[h!]
	\begin{center}		
		\includegraphics[scale=0.35]{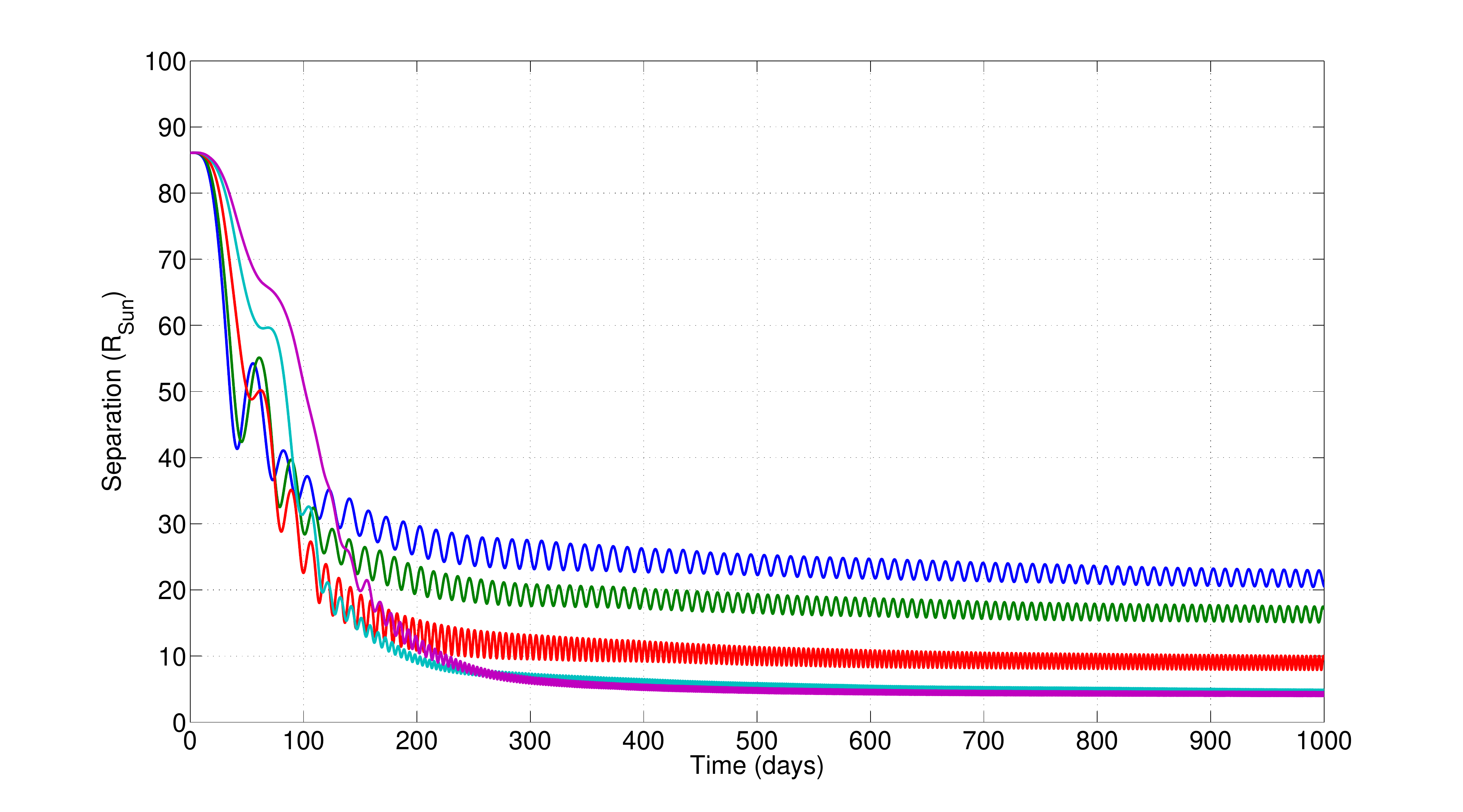}
	\caption{Separation between the primary core and the companion as a function of time for the $256^3$ {\it Enzo} simulations. The companion masses are 0.9 (blue), 0.6 (green), 0.3 (red), 0.15 (cyan) and 0.1 (purple)~\msun.
	\label{fig:orbitstotal}
	}
	\end{center}
\end{figure}

\begin{figure}[h!]
	\begin{center}
		\includegraphics[scale=0.35]{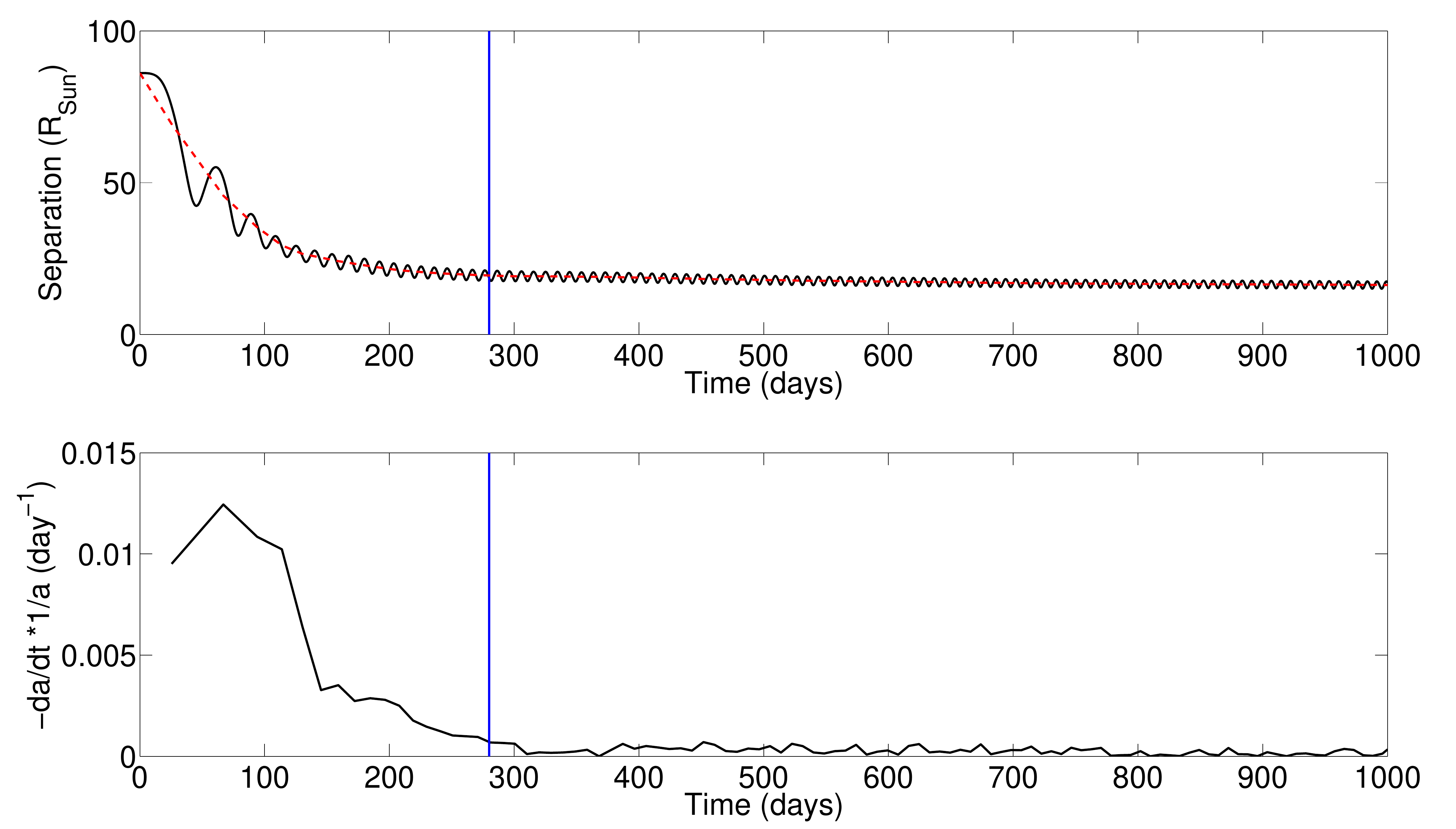}
	\caption{Evolution of the separation (top) and of the orbital decay (bottom) for Enzo7. The orbital decay is computed using orbital separations averaged over each cycle (red dashed line). The blue vertical line shows the time when we define the end of the rapid infall phase.
	\label{fig:decay}
	}
	\end{center}
\end{figure}

As explained in \S\ref{sec:simus}, the companion is placed at the surface of the primary. Thus, the primary extends beyond its Roche lobe and unstable mass transfer starts immediately. The companion, surrounded by stellar matter, exchanges momentum and  energy with this gas through drag. The orbital separation shrinks on a dynamical timescale and its evolution for the $256^3$ {\it Enzo} simulations is shown in Fig.~\ref{fig:orbitstotal}. Although the orbit is initially circular, it quickly develops eccentricity due to the geometry of the gas ejection. In order to define quantitatively the end of the rapid infall phase and the final orbital separation {\it ad hoc}, we consider the evolution of the orbital decay (Fig.~\ref{fig:decay}). As expected, the orbital decay is initially quite high ($\sim 0.01$~day$^{-1}$), decreases as less gas is available for the companion to exchange energy with, and eventually reaches a plateau. We decide to define the end of the rapid infall phase to be at the start of this plateau, which occurs at about 280~days for the 0.6~\msun \ companion (Fig.~\ref{fig:decay}). All the simulations show the same trend and the lighter the companion, the deeper it falls and the longer it needs to reach its final orbital separation. The duration of the rapid infall phase is 260, 280, 280, 300 and 340~days, for the 0.9, 0.6, 0.3, 0.15 and 0.1~\msun \ companion, respectively.

\begin{figure}[h!]
	\begin{center}	
		\includegraphics[scale=0.33]{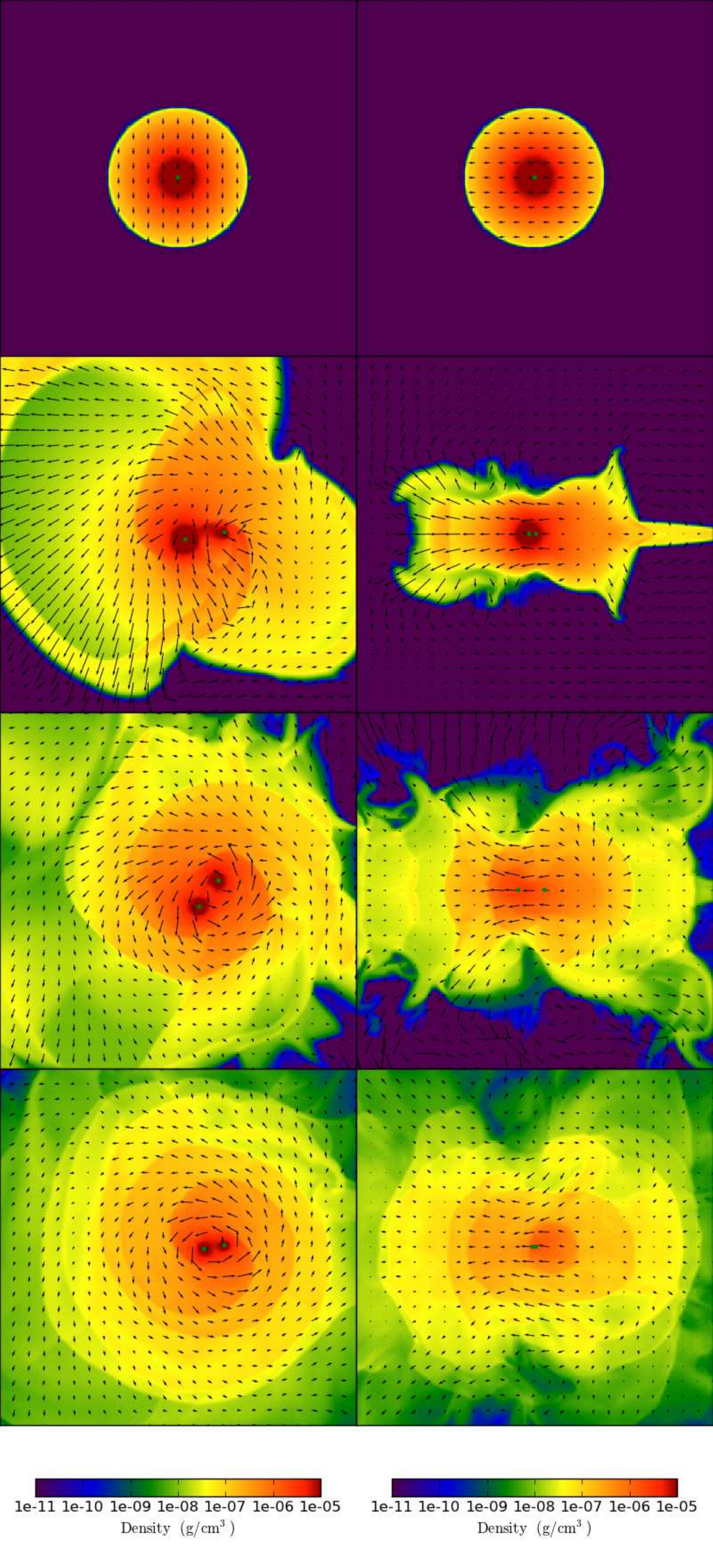}
	\caption{Density slices in the orbital plane (left) and in the perpendicular plane (right) at 0, 50, 85 and 130 days (from top to bottom) for the Enzo7 simulation. The scale used for the velocity vector field is the same on each frame and is such that the velocity shown on the top panel equals the initial orbital velocity of the primary ($\sim$ 23 \kms).
	\label{fig:cuts1}
	}
	\end{center}
\end{figure}

\begin{figure}[h!]
	\begin{center}
		\includegraphics[scale=0.5]{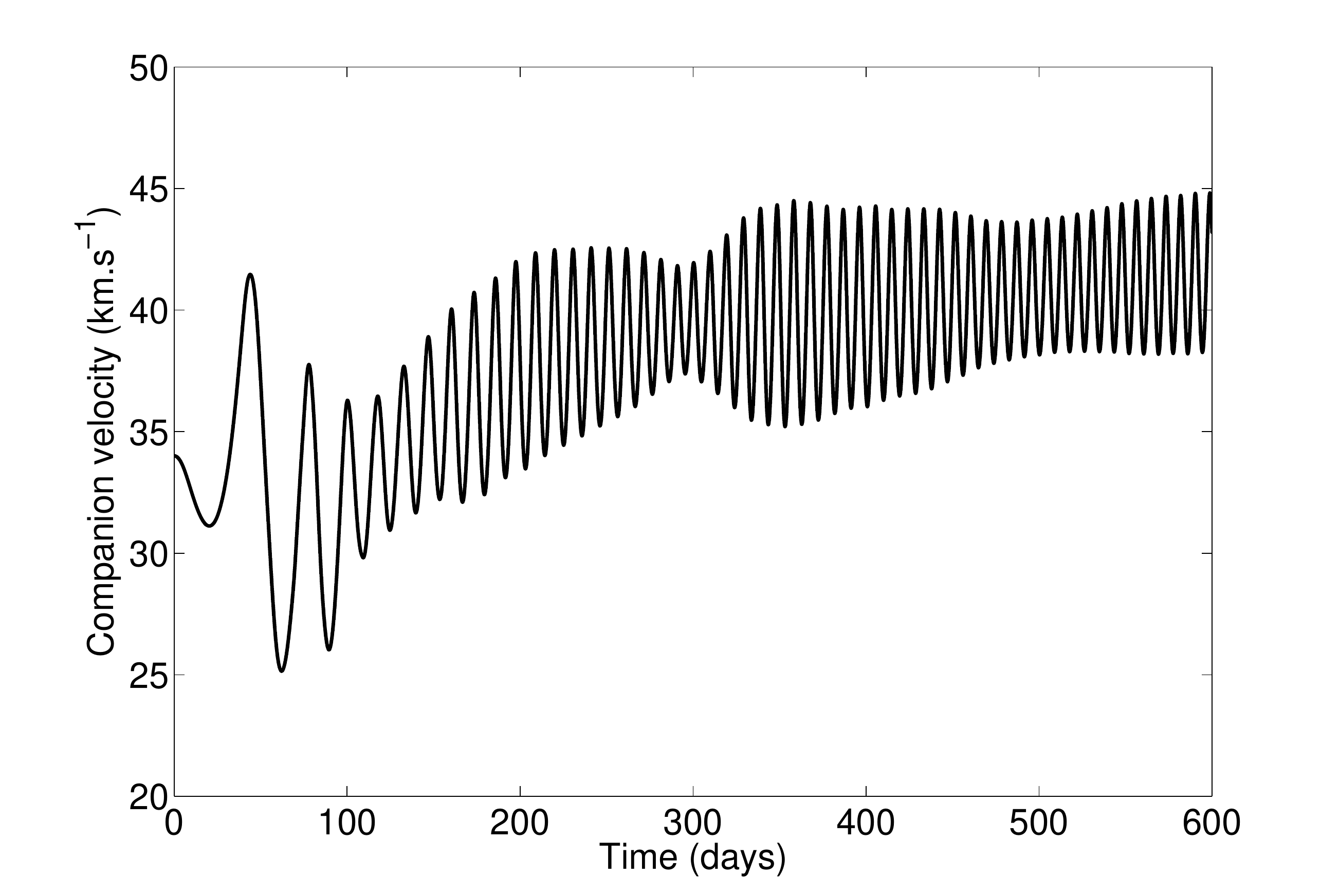}
	\caption{Evolution of the companion velocity for the Enzo7 simulation.
	\label{fig:velocity}
	}
	\end{center}
\end{figure}

As orbital energy is transferred to the envelope, the latter is ejected, initially in the orbital plane; at later phases there is an almost equal distribution of matter into the polar direction as well (Fig.~\ref{fig:cuts1}). Overall, almost 90\% of the envelope is ejected within an angle of 30$^\circ$ on each side of the equatorial plane. We compare the orbital velocity of the companion (Fig.~\ref{fig:velocity}) with the local sound speed of the gas (Fig.~\ref{fig:profiles}, bottom left panel). The former does not exceed 50 \kms \ while the highest sound speed encountered is about 60 \kms. The companion moves only slightly above or below the local sound speed. We therefore conclude that the SPH noise could not significantly influence the solution. Also, since the motion of the companion is not highly supersonic, the shocks are not strong and we can use {\it Enzo} with the faster Zeus solver.

Unlike the SPH computational domain, the {\it Enzo} grid is spatially limited. Thus, the evolution of the gas that leaves the grid cannot be followed. Therefore, we use the SPH2 simulation to study the global evolution of the angular momentum and the energy of the system. 

\begin{figure}[h!]
	\begin{center}
		\includegraphics[scale=0.34]{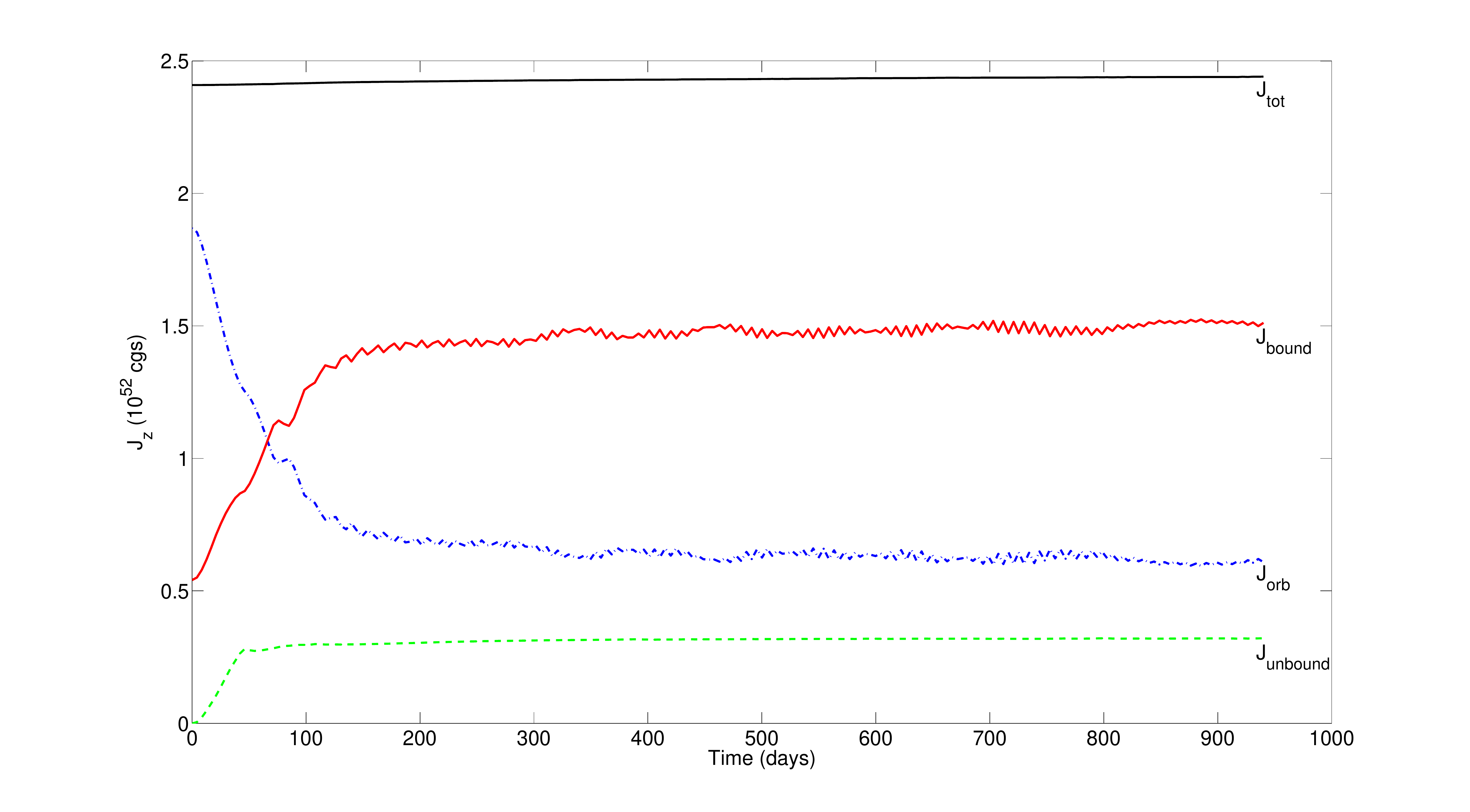}
	\caption{Evolution of the z-component of the total angular momentum ($J_{\rm tot}$), the angular momentum of the core and the companion ($J_{\rm orb}$), the angular momentum of the bound mass ($J_{\rm bound}$) and the angular momentum of the unbound mass ($J_{\rm unbound}$) for the SPH2 simulation.
	\label{fig:AM}
	}
	\end{center}
\end{figure}

We compute the angular momentum using the center of mass of the SPH particles as the center of reference. As shown in Fig.~\ref{fig:AM} for the 0.6~\msun \ companion case, the total angular momentum of the system is conserved to less than 1\%. Since the ejection of the gas is asymmetric, the center of reference is eventually located outside the orbit. Consequently, studying the orbital components individually is irrelevant as the sign of each component changes during a single orbit. Therefore, we study their sum $J_{\rm orb}$ instead. During the first 50~days, angular momentum from the orbit almost equally spins up the envelope and unbinds mass from the outer layers. Later on, no more additional mass gets unbound (see \S~\ref{sec:envelope}) and the angular momentum lost from the orbit spins up the bound envelope only. Since the unbound mass is located at large distances from the primary's core, there is no more exchange of angular momentum between the unbound mass and the rest of the system. After~$\sim$~150~days, there is no more angular momentum exchange in the system. The primary's core and the companion --- which are the main contributors to the calculation of the centre of mass --- switch positions twice per orbit, which leads to small periodic motions of the center of mass. These periodic displacements are the causes for the small angular momentum fluctuations of the orbital components and the bound mass occurring after 100~days.

\begin{figure}[h!]
	\begin{center}
		\includegraphics[scale=0.42]{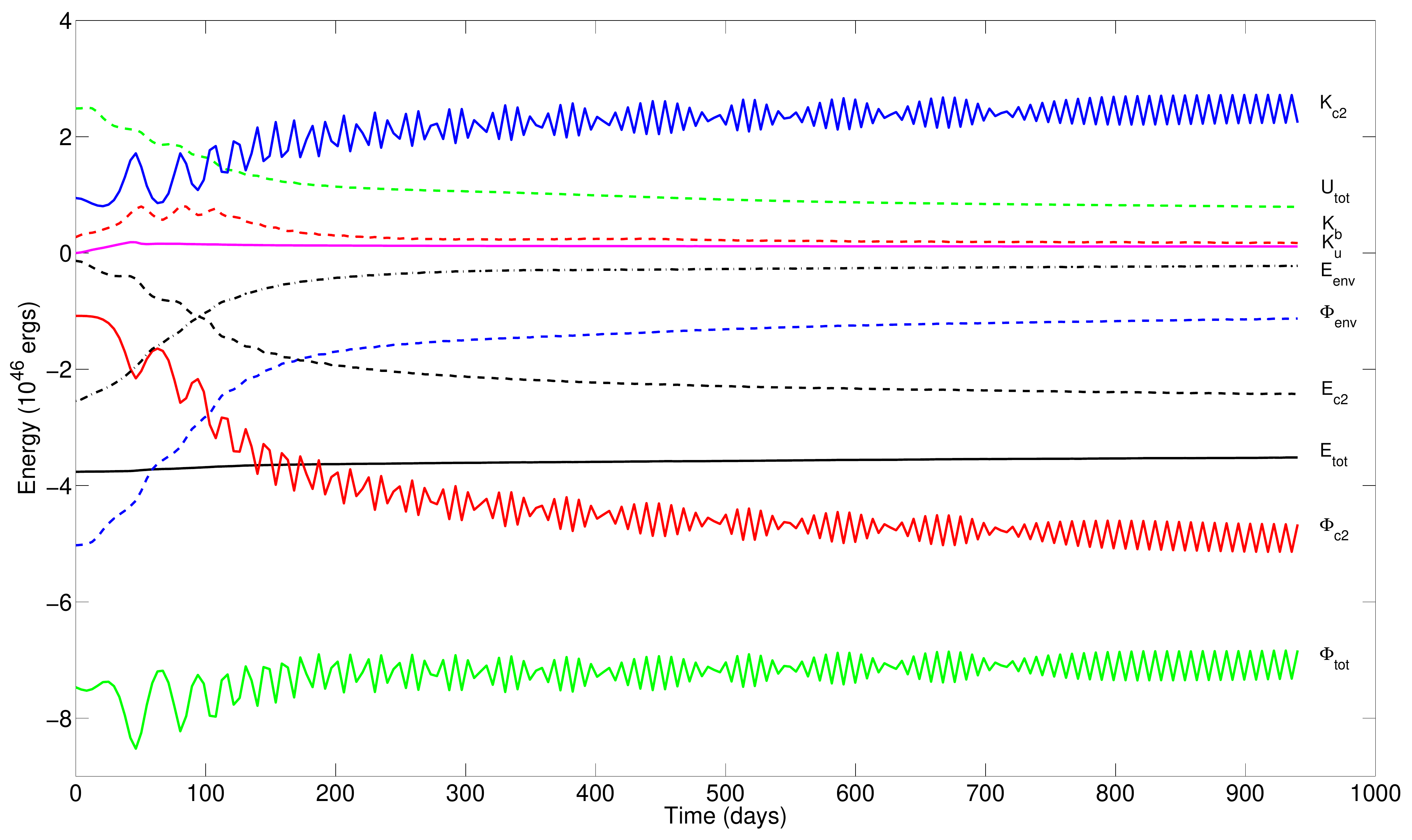}
	\caption{Energy components\ for the SPH2 simulations. Plotted are the total energy ($E_{\rm tot}$), the total gravitational potential energy $\Phi_{\rm tot}$, the internal energy of the system ($U_{\rm tot}$), the gravitational potential energy of the envelope ($\Phi_{\rm env}$), the gravitational potential energy from the core-companion interaction ($\Phi_{\rm c2}$), the kinetic energy of the core and the companion ($K_{\rm c2}$), the kinetic energy of the bound mass ($K_{\rm b}$), the kinetic energy of the unbound mass ($K_{\rm u}$), the orbital energy of the core-companion system ($E_{\rm c2}$) and the total energy of the envelope ($E_{\rm env} \equiv \Phi_{\rm env} + U_{\rm tot} + K_{\rm b}$). The beat frequency seen on $K_{c2}$ and $\Phi_{c2}$ are due to the non-synchronization between the orbital period and the data dumping frequency.
	\label{fig:energy}
	}
	\end{center}
\end{figure}

We plot the various energy components in Fig.~\ref{fig:energy}. We start by explaining the different components of potential, thermal and kinetic energies represent and how they are computed. Among numerous other attributes, each particle $i$ possesses a specific gravitational potential energy $\phi_i$, a specific thermal energy $u_i$ and a specific macroscopic kinetic energy $k_i$. By definition,

\begin{equation}
	\phi_i = \sum_{\rm j \ particles, \ j \neq i} -G \frac{M_j^{\rm grav}}{r_{ij}}
\end{equation}

\noindent where G is the gravitational constant, $M_j^{\rm grav}$ is the gravitational mass of particle $j$ and $r_{ij}$ is the distance between particles $i$ and $j$. We compute these different components using the gravitational mass of the particle for the gravitational potential energy and the macroscopic kinetic energy, and the SPH mass for the thermal energy (see \S\ref{sec:simus}):

\begin{eqnarray}
	\Phi_i = M_i^{\rm grav} \phi_i \\
	K_i = M_i^{\rm grav} k_i \\
	U_i = M_i^{\rm sph} u_i
\end{eqnarray}

\noindent where $M_i^{\rm sph}$ is the SPH mass of particle $i$ used to compute its acceleration due to pressure. We recall that both masses are identical for all particles except the primary's core and the secondary. Thus, the total gravitational potential energy of the system is

\begin{equation}
	\Phi_{\rm tot} = \frac{1}{2} \sum_{\rm i \ particles} \Phi_i
\end{equation}

Finally, we subtract the contribution of the secondary from the total potential energy in order to calculate the binding energy of the envelope:

\begin{equation}
	\Phi_{\rm env} = \Phi_{\rm tot} - M_2^{\rm grav} \phi_2
\end{equation}

\noindent where the subscript ``2" stands for the secondary.

\begin{figure}[h!]
	\begin{center}
		\includegraphics[scale=0.5]{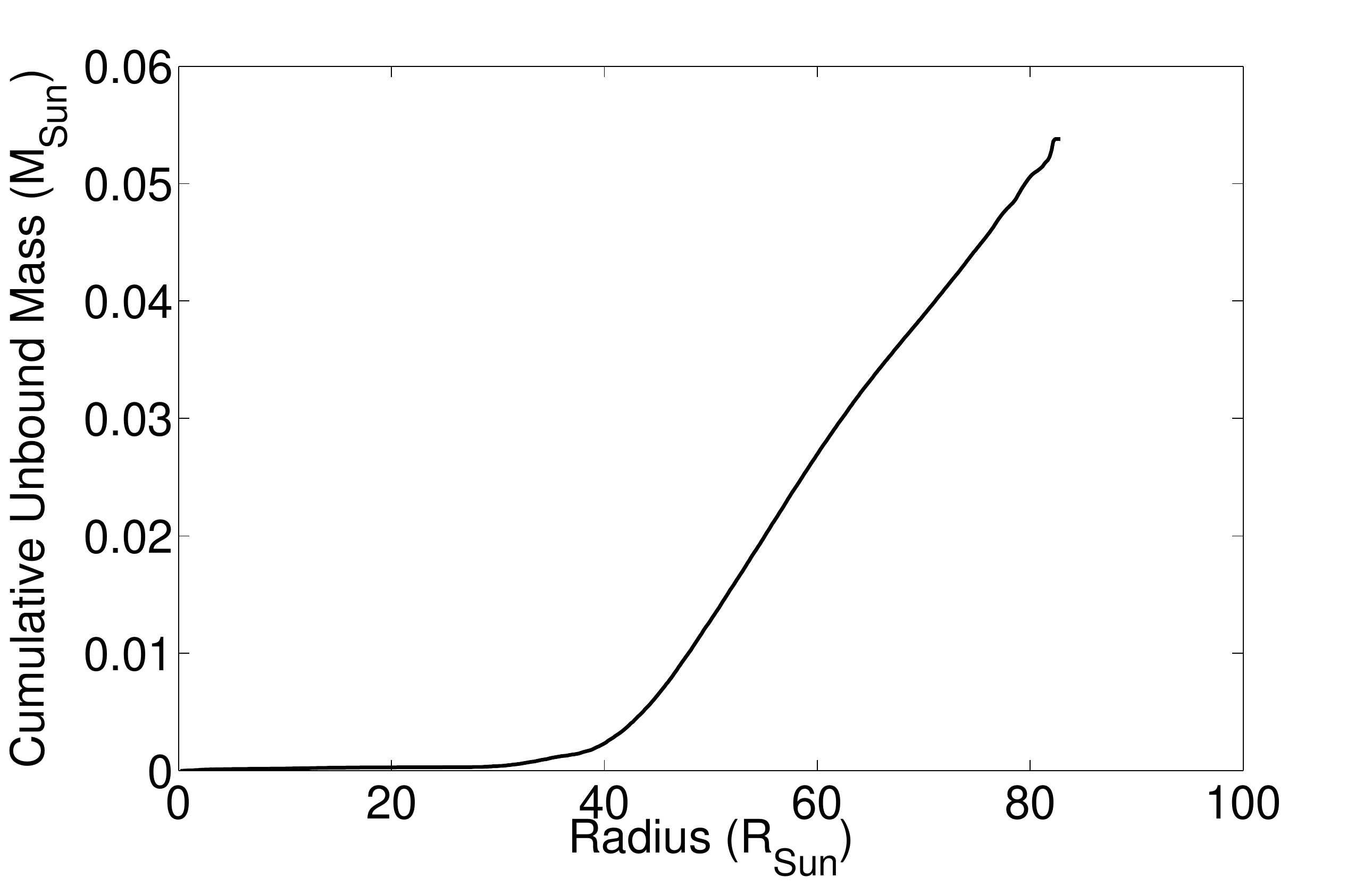}
	\caption{Initial distribution within the envelope of the mass that will eventually get unbound for SPH2. 
	\label{fig:unboundprofile}
	}
	\end{center}
\end{figure}

During the first 200 days when most of the in-spiral happens, the total internal energy of the system decreases by more than a factor of two: the envelope expands and therefore cools. The energy released is transferred mostly into macroscopic kinetic energy of the gas: the envelope is lifted up, accelerated and the outermost part of the envelope becomes unbound in the first 50 days. At later times, more energy is transferred from the orbit to the envelope but no more material becomes unbound. One can easily note in Fig.~\ref{fig:energy} how the variations of the orbital energy of the core-secondary system and of the total energy of the envelope balance each other. The total energy of the envelope remains negative throughout the simulation. We follow the evolution of the unbound particles and determine their initial position in the envelope. Fig.~\ref{fig:unboundprofile} shows the cumulative mass of the particles that will eventually get unbound as a function of their initial distance from the core. It confirms that the unbound mass was initially located in the outer part of the envelope and that almost all gas located initially closer than 40~\rsun \ from the primary's core remains bound at the end of the simulations.

\subsection{Code comparison}

\begin{figure}[h!]
	\begin{center}
		\includegraphics[scale=0.45]{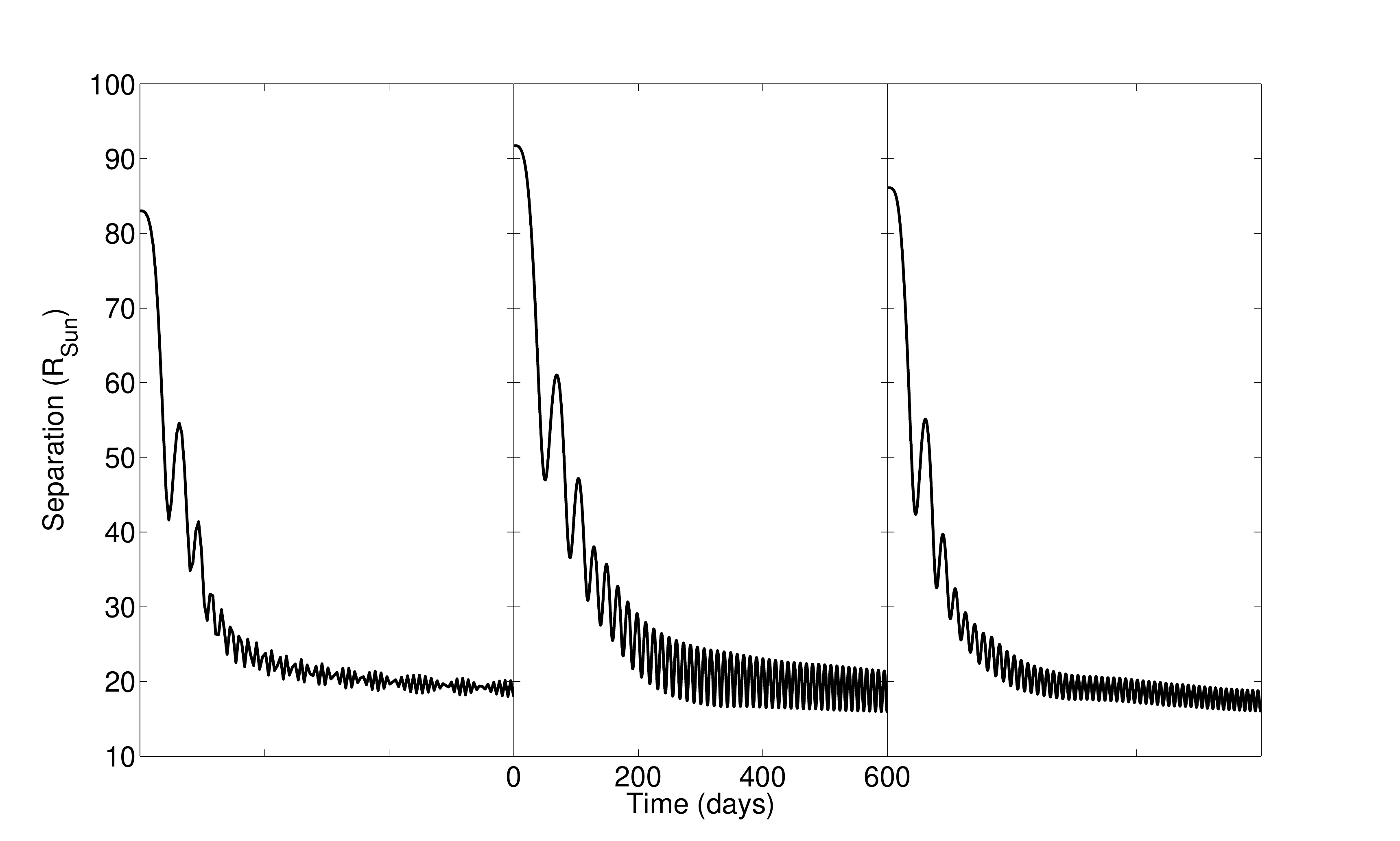}
	\caption{Separation between the core of the primary and the 0.6~\msun \ companion as a function of time for the SPH2 (left), Enzo2 (middle) and Enzo7 (right) simulations. Again, the beat frequency seen in the SPH simulation is due to the non-synchronization between the orbital period and the dumping frequency.
	\label{fig:orbits}
	}
	\end{center}
\end{figure}

\begin{figure}[h!]
	\begin{center}
		\includegraphics[scale=0.45]{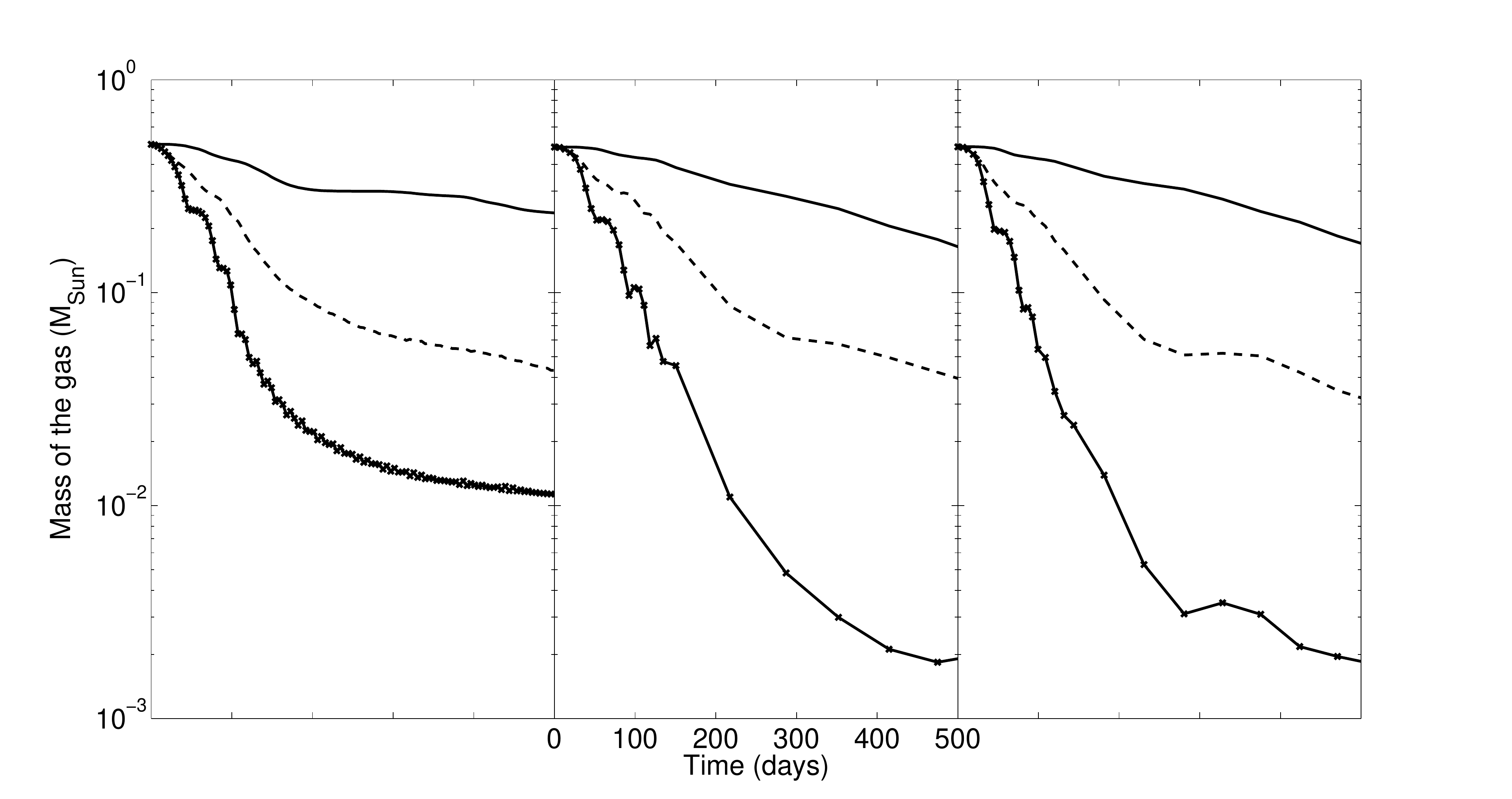}
	\caption{Each panel shows the mass within the equivalent {\it Enzo} grid (plain), the inital volume of the primary (dash) and the orbit (cross-solid) as a function of time for the SPH2 (left), Enzo2 (middle) and Enzo7 (right) simulations.
	\label{fig:mass}
	}
	\end{center}
\end{figure}

\begin{figure}[h!]
	\begin{center}
		\includegraphics[scale=0.3]{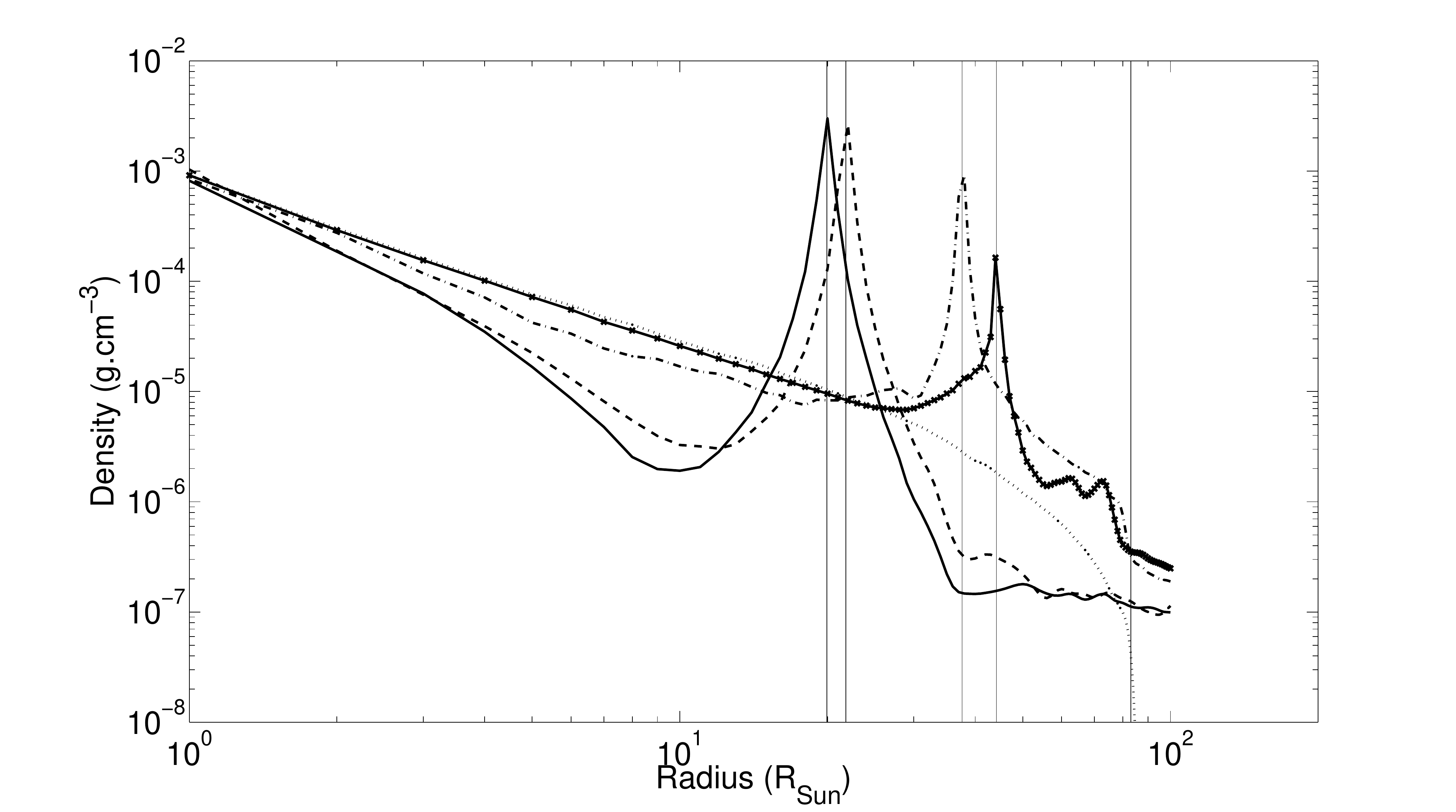}
		\includegraphics[scale=0.3]{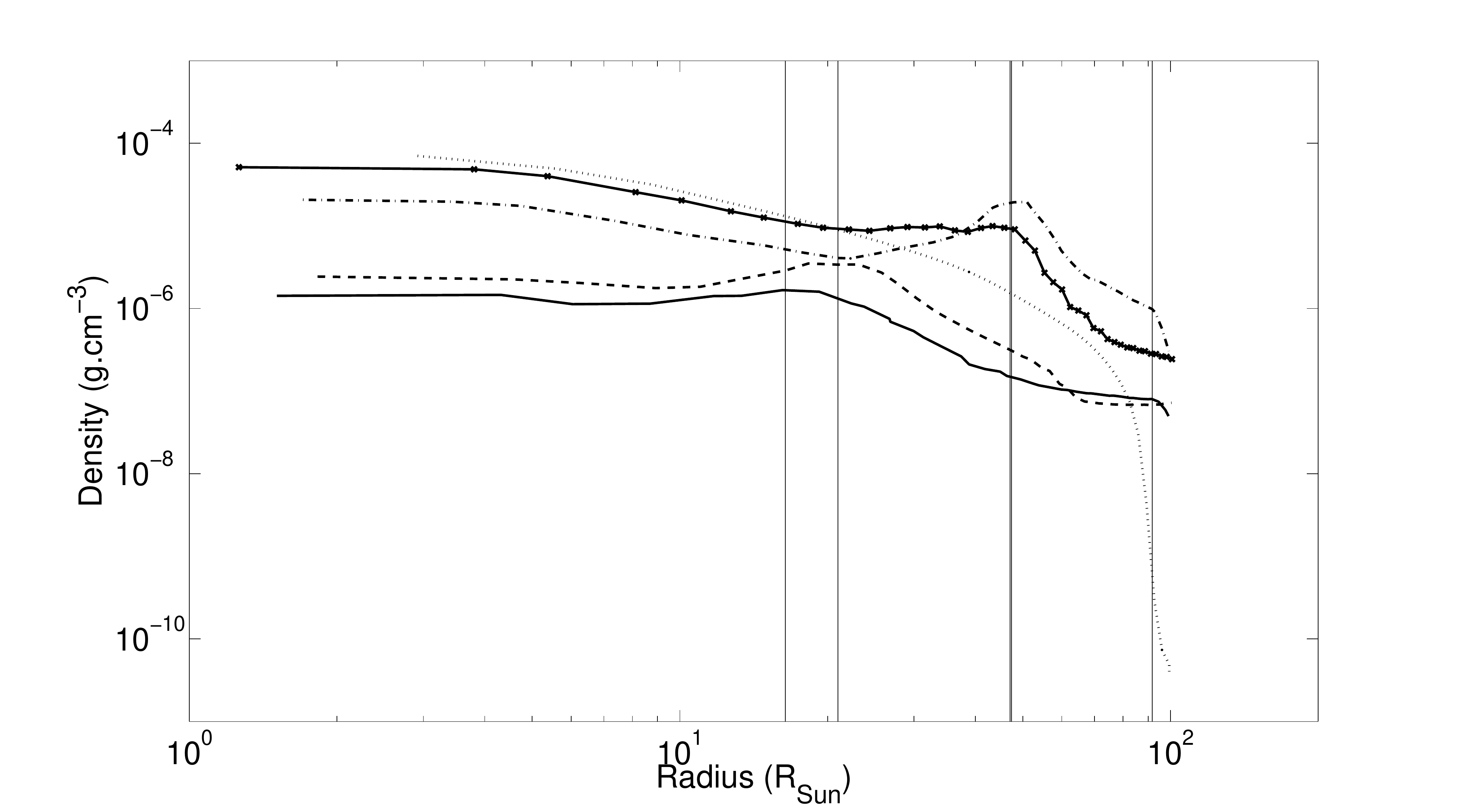}
		\includegraphics[scale=0.3]{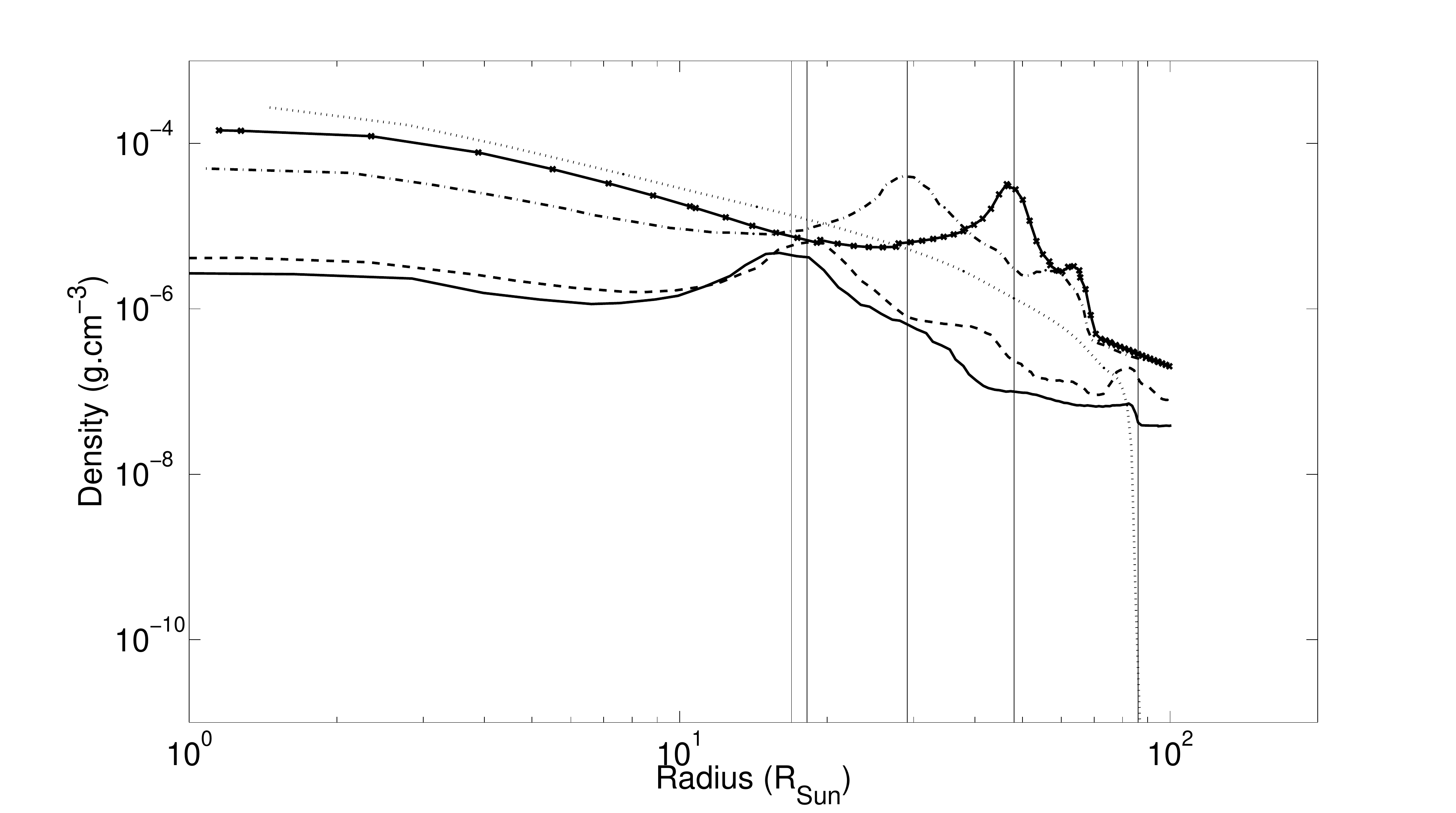}
	\caption{Density profiles along the line joining the core and the 0.6~\msun \ companion at 0 (dotted line), 50 (dash-cross line), 100 (dash-dot line), 300 (dashed line) and 500 (solid line) days for SPH2 (top), Enzo2 (middle) and Enzo7 (bottom). The vertical lines show the position of the companion.
	\label{fig:densityline}
	}
	\end{center}
\end{figure}

The fact that a code solves the equations in an accurate and precise way in a particular situation does not necessary mean it will do so in another regime. Thus, a direct comparison of simulations of the CE interaction using two different numerical methods is a good solution for testing the ability of the two methods to model this problem. One can see in Fig.~\ref{fig:orbits} and Table~\ref{tab:runs} that for each binary system, the final separations in the {\it Enzo} simulations are very close to those obtained with the equivalent {\it SNSPH} simulations. We may then compare the mass evolution of the material in the volume defined by the {\it Enzo} grid, the matter within the initial volume of the primary, and within the current separation. For the 0.6~\msun \ companion (Fig.~\ref{fig:mass}), both the mass within the {\it Enzo} grid and the mass within the initial volume of the progenitor agree well between the {\it Enzo} and the {\it SNSPH} runs. For the mass within the orbit we notice a difference of $\sim 10^{-2}$~\msun \ between the {\it Enzo} and the {\it SNSPH} runs. This difference is large compared with the mass of a SPH particle ($\sim 10^{-6}$~\msun) and is due to how accurately accretion of the gas by the core and the companion is resolved by the two codes. We have plotted, in Fig.~\ref{fig:densityline}, density profiles at different times along the line joining the primary core and the secondary, for the three simulations with the 0.9~\msun \ companion. Accretion onto the secondary is better resolved in the SPH simulations in which the maximum density of the matter accreted by the companion is about $10^{-3}~{\rm g\,cm^{-3}}$. This maximum value depends on the resolution of the runs. In the single-grid {\it Enzo} runs, accretion is poorly resolved due to the low number of cells resolving the local region around each particle. Although mass is still accreted around the particles, it eventually becomes dispersed. For the SPH runs, around 60 particles interact within a smoothing length so the accretion zone is well resolved. On the other hand, the cell width of the {\it Enzo} $256^3$ runs is about 1.6 \rsun \ so the accretion zone cannot be resolved although it is still better than for the {\it Enzo} $128^3$ simulations as can be seen from comparing the different density profiles at 50 days (Fig.~\ref{fig:densityline}). However, the accurate simulation of accretion onto the secondary is not crucial for the global evolution of the system: as we mentioned earlier, the evolution is not driven by accretion but by drag forces. Although the density of the matter accreted by the companion differs by up to 3 orders of magnitudes between the two methods, the accreted mass is negligible compared with the companion mass and the final orbital separations are very similar.

\cite{RickerTaam2008} used the FLASH code \citep{FlashCode} to study the CE evolution of a binary system consisting of a 1.05~\msun \ RGB star having a 0.36~\msun \ core and a 0.6~\msun \ companion. Their implementation is somewhat different from ours since they treat the red giant core and the companion as spherical clouds of particles. In spite of those differences, their progenitor is almost identical to ours and they find a final separation of 20 \rsun \ which falls within the range of the results given by our simulations SPH2, Enzo2 and Enzo7. Moreover, one can see in Fig.~\ref{fig:velocity} that for the 0.6~\msun \ companion, the velocity of the companion stays below 50 \kms \ and therefore, the gas flows are subsonic except in the outer layers. This conclusion was also reached by \cite{RickerTaam2008}.

\subsection{The impact of initial conditions}

\begin{figure}[h!]
	\begin{center}
		\includegraphics[scale=0.4]{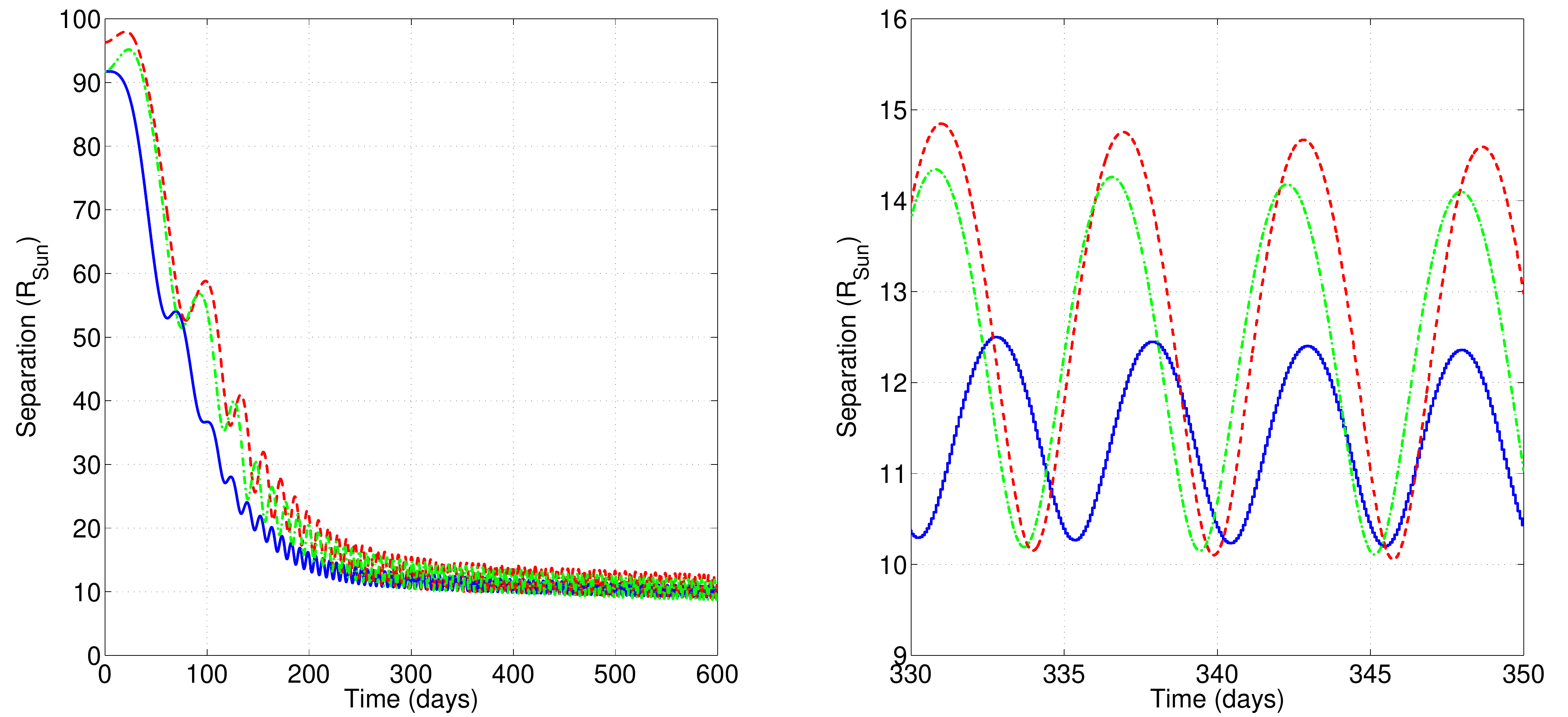}
	\caption{Left: separation between the core of the primary and the companion as a function of time for the Enzo3 (solid blue), Enzo11 (dashed red) and Enzo12 (dash-dot green) simulations. Right: a detail of the comparison from the left panel at $\sim 340$ days.
	\label{fig:sensi}
	}
	\end{center}
\end{figure}

In order to determine the sensitivity of the final state of the system to the initial parameters, we start with the Enzo3 simulation and increase by 5\% either the initial velocity of the secondary (Enzo11) or the initial separation between the two particles (Enzo12), which correspond to initial eccentricities of 0.10 (Enzo11) and 0.05 (Enzo12). The evolution of the separation for those three simulations is compared in Fig.~\ref{fig:sensi}. For Enzo11 and Enzo12, the ratio of the initial velocity of the companion to the velocity required for a circular orbit is higher than one ($v_0/v_{circ} > 1$), so the separation must first increase. The larger the orbital separation, the more delayed the rapid infall phase is and the later the system reaches its final separation. The final separations for Enzo3, Enzo11 and Enzo12 are 11.7, 12.0 and 12.2 \rsun , respectively, and the final eccentricities are 0.09, 0.17 and 0.18, respectively. As expected, the companion that moves outwards the farthest initially, sinks into the envelope with a higher orbital decay velocity. Therefore, it attains a more eccentric orbit and completes fewer revolutions around the primary core (Fig.~\ref{fig:sensi}). However, the standard deviation of the final separation between the three simulations ($\sigma \sim 0.2$~\rsun) is more than 10 times smaller than the width of a cell. Consequently, we conclude that the final results are quite insensitive to the initial conditions at the level tested.

\subsection{Gravitational vs Hydrodynamic drag}
\label{subsec:drag}

The drag exerted on the companion has two components: gravitational and hydrodynamical. The former is due to gravitational forces from matter flowing past the companion and colliding with its wake \citep{BondiHoyle1944, Iben1993}, while the latter is due to ram pressure forces on the companion. The hydrodynamical contribution can be estimated as:

\begin{equation}
	F_{\rm hydro} \sim \rho \mathbf{v}_2^{2} \times  \pi R_2^2
	\label{eq:hydro_drag}
\end{equation}

\noindent where $R_2$ is the radius of the secondary, $\mathbf{v}_2$ is the relative velocity between the secondary and the envelope, and we have taken the coefficient of drag to be unity for simplicity. In a similar manner, the gravitational drag is approximated by \citep{Iben1993}:

\begin{equation}
	F_{\rm grav} \sim  \rho \mathbf{v}_2^{2} \times \pi R_A^2
	\label{eq:grav_drag}
\end{equation}

\noindent where the accretion radius $R_A$ is defined as:

\begin{equation}
	R_A = \frac{2GM_2}{\mathbf{v}_2^2 + c_s^2}
\end{equation}

\noindent where $c_s$ is the sound speed of the medium. Choosing $|\mathbf{v}| = 2 c_s = 80$~\kms \ with an 0.6 \msun \ companion yields $R_A \sim 30$~\rsun. Assuming $R_2 \sim 1$~\rsun, we conclude that the hydrodynamical drag is of the order of almost 1\,000 times smaller than the gravitational drag, thus negligible. 

This conclusion is also confirmed by the outcomes of our simulations. Indeed, the primary's core and the companion are treated as point masses and are not pressure sources, except for the primary's core in the {\it SNSPH} simulations. Instead of being caused by the finite size of the particles, hydrodynamical drag in the models is thus due to the matter accreted around them. We pointed out earlier that the accuracy with which accretion was treated was different between the two different models because of the different finest resolutions and softenings used: accretion is poorly modeled in the {\it Enzo} simulations whereas in the {\it SNSPH} simulations, the companion builds up a sphere of accreted matter about a few \rsun \ wide around itself (Fig.~\ref{fig:densityline}). This should lead to differences in the magnitude of hydrodynamic drag forces. Nevertheless, the consistency of the results suggests that the hydrodynamic drag is unimportant in the evolution of the system, confirming the results of \cite{RickerTaam2008}.

\section{Discussion}
\label{sec:discussion}

\subsection{Comparison of simulations and observations}
\label{subsec:alphacomp}

We now compare the numerical results with a sample of 61 observed post-CE systems listed in \cite{Zorotovic2010} and \cite{AlphaPaper2011}. 

\subsubsection{Final separations}

\begin{figure}[h!]
	\begin{center}
		\includegraphics[scale=0.45]{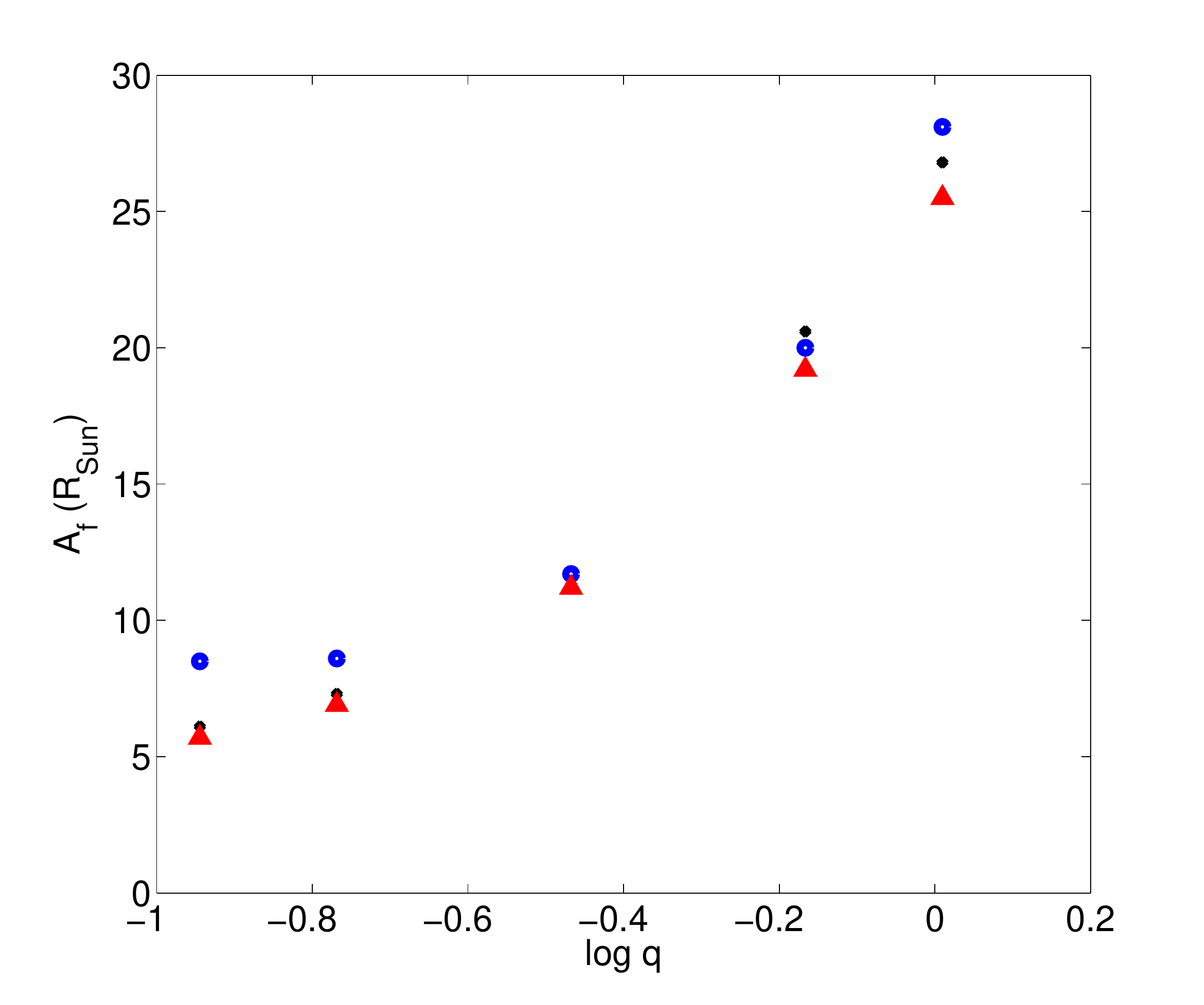}
	\caption{Final separations as a function of the mass ratio $q$ for the {\it SNSPH} (black cross), {\it Enzo} $128^3$ (blue circle) and {\it Enzo} $256^3$ (red triangle) simulations.
	\label{fig:sepsims}
	}
	\end{center}
\end{figure}

For a given companion mass (or alternatively mass ratio $q$) we obtain 3 values for the final separation $A_f$, one for each simulation carried out with that companion mass (Table \ref{tab:runs} and Fig.~\ref{fig:sepsims}). One can distinguish between these values at high $q$ ($q \geq 0.34$), which correspond to ``heavy'' companions ($M_2 \geq 0.3$~\msun), and the ones at low $q$ ($q < 0.34$) corresponding to ``light'' companions ($M_2 < 0.3$~\msun). At high $q$, the values of $A_f$ are very similar and the standard deviation is more than 20 times smaller than the average value of $A_f$. At low $q$, the companion sinks deeper and as a consequence, the resolution used in the $128^3$ {\it Enzo} simulations is not sufficient. However, as one increases the resolution to $256^3$ cells, the final separations converge to the solutions given by the {\it SNSPH} simulations.

\begin{figure}[h!]
	\begin{center}
		\includegraphics[scale=0.5]{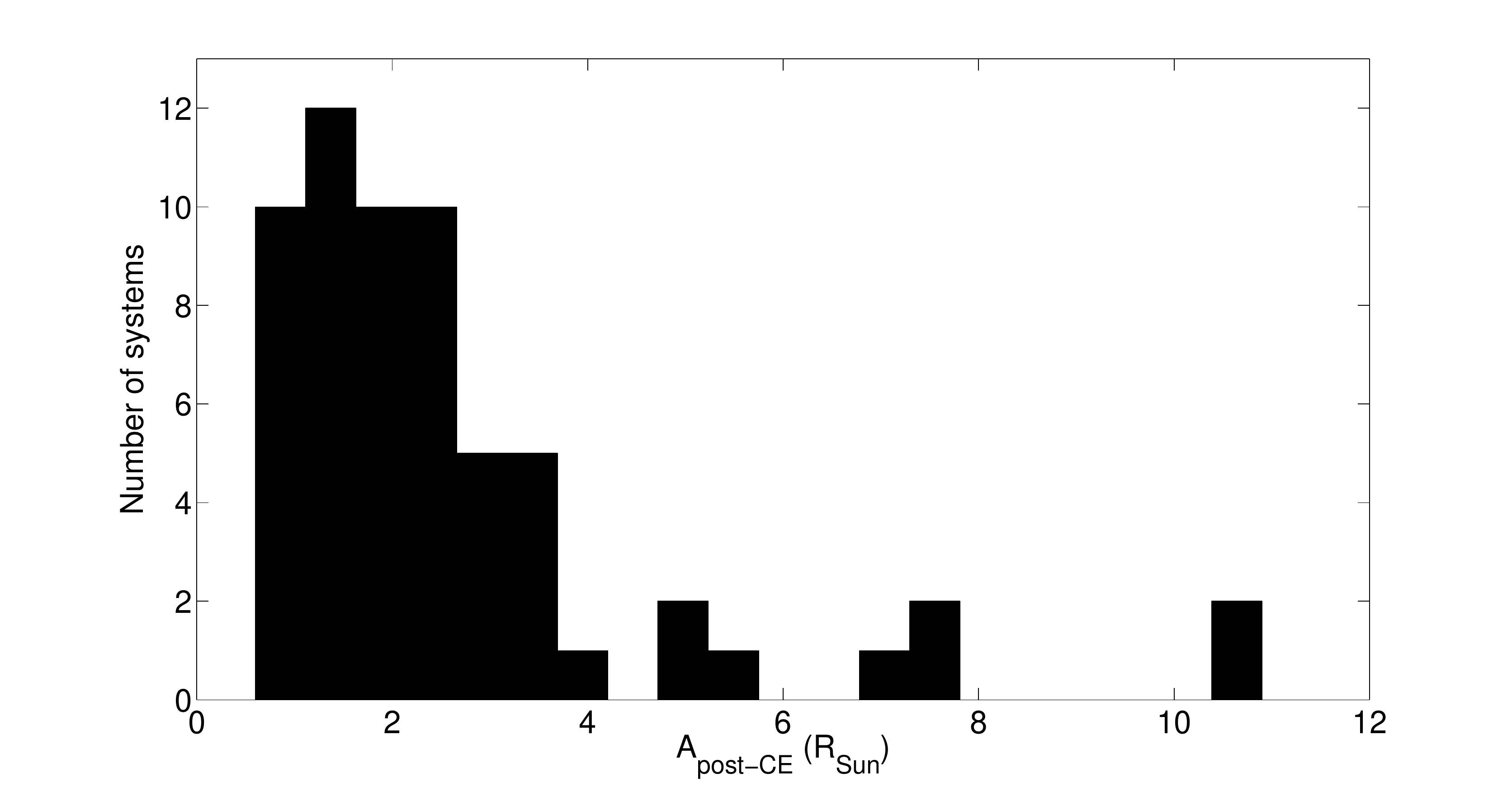}
	\caption{Distribution of post-CE systems as a function of their observed orbital separation from \cite{Zorotovic2010} and \cite{AlphaPaper2011}.
	\label{fig:histo}
	}
	\end{center}
\end{figure}

\begin{figure}[h!]
	\begin{center}
		\includegraphics[scale=0.39]{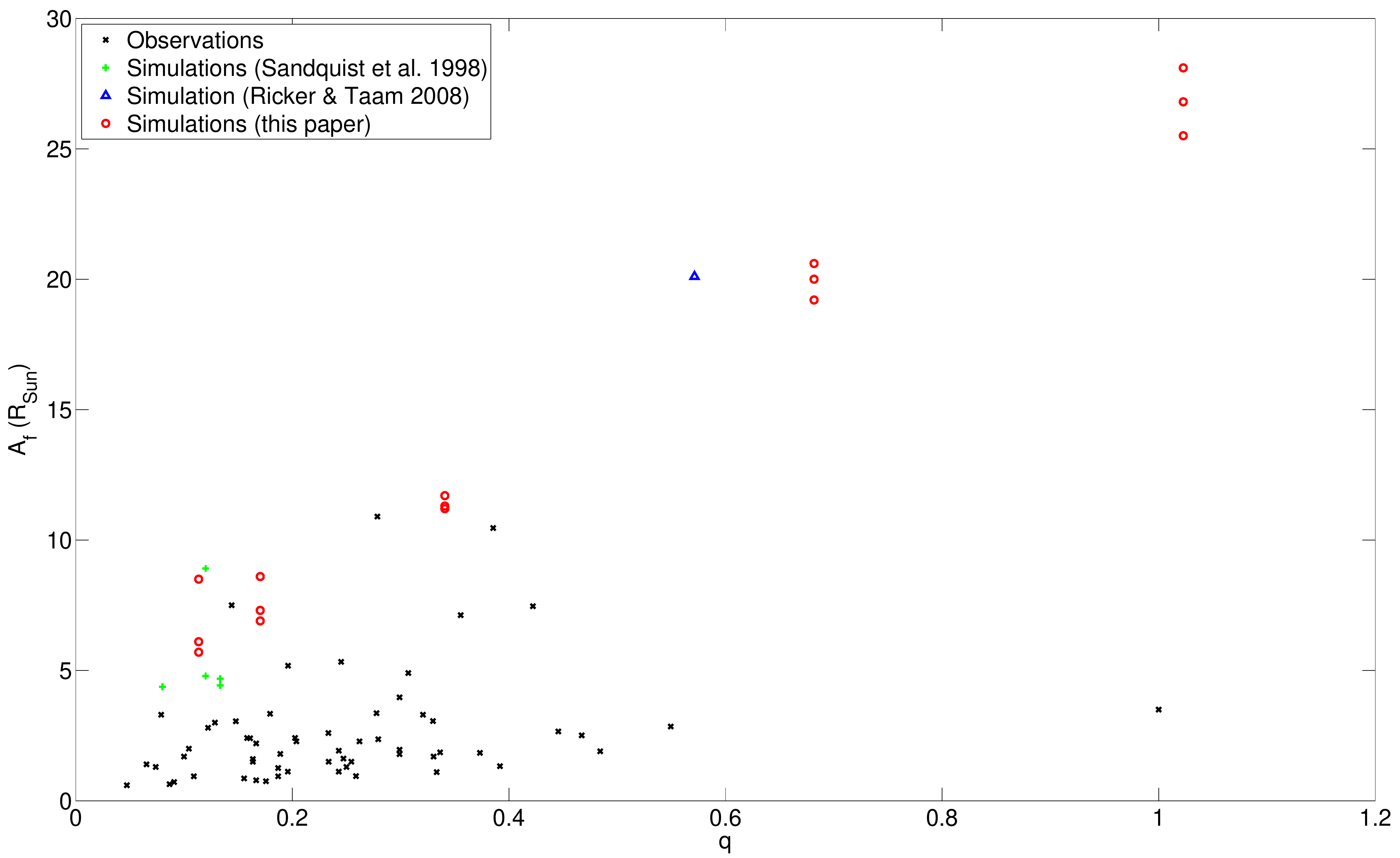}
	\caption{Comparison between the orbital separations of observed post-CE systems (black dot) and the final separations reached at the end of the simulations (red circle), as well as the ones by \cite{Sandquist1998} (green circles) and by \cite{RickerTaam2008} (blue triangle).
	\label{fig:sepobs}
	}
	\end{center}
\end{figure}

Fig.~\ref{fig:histo} shows the distribution of orbital separations reached by the 61 post-CE systems. For all these systems, there has been no substantial orbital shrinkage due to phenomena such as magnetic braking or radiation of gravitational waves \citep[see discussion in][]{Schreiber2003}. Although they cover a significant range in secondary masses, going from a 1.1~\msun \ MS star down to a 0.05~\msun \ brown dwarf, all of them have separations smaller than 11~\rsun. Furthermore, 87\% of those systems have separations smaller than 4~\rsun, which is smaller than any value obtained in our simulations. This is even more obviously shown in Fig.~\ref{fig:sepobs}, where the final separations for simulations presented here and in the literature are compared to the orbital separations of the observed post-CE systems. Although a couple of observed systems have $q \geq 0.5$, one clearly sees that the simulations with $M_2 = 0.9$ and 0.6~\msun \ leave the companion far out. Systems with lower mass companions ($M_2 \leq 0.3$~\msun) have by and large lower orbital separations than in our simulations. Simulations of \cite{Sandquist1998} and \cite{RickerTaam2008} shown in Fig.~\ref{fig:sepobs} give results consistent with ours. All these numerical simulations suggest that the separations between the secondary and the primary's remnant at the end of the simulated rapid infall phase are too large to explain the orbital separation of the currently observed post-CE systems. This suggests that further evolution of the orbital separation must occur during the phase immediately following the rapid infall phase. We discuss this point further in \S\ref{subsec:reasons}.

\subsubsection{The state of the envelope at the end of the simulations}
\label{sec:envelope}

\begin{figure}[h!]
	\begin{center}
		\includegraphics[scale=0.45]{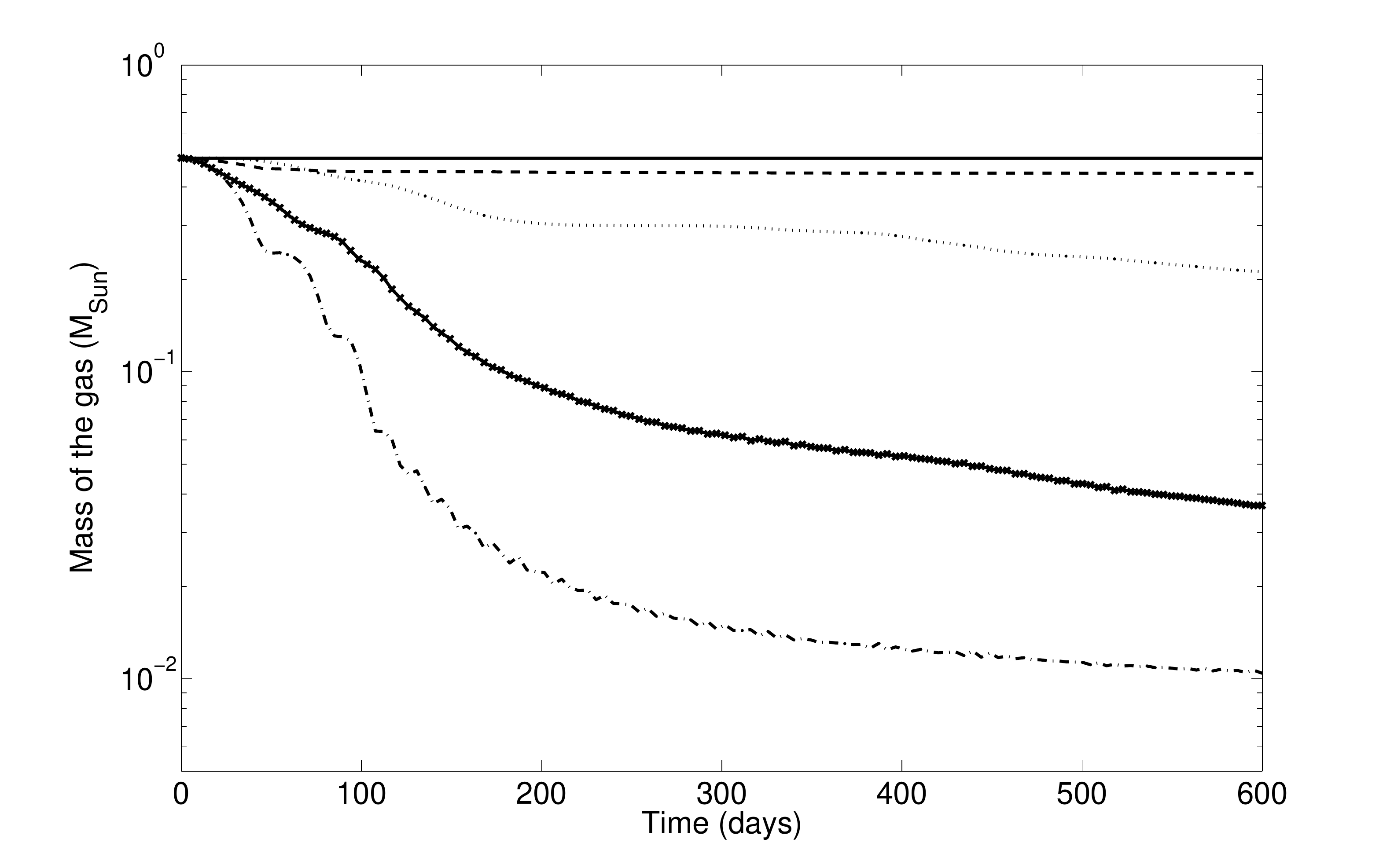}
	\caption{Evolution of the total mass (solid line), the bound mass (dashed line), the mass within the volume of the {\it Enzo} grid (dotted line), the mass within the initial volume of the primary (dash-cross line) and the mass within the orbital separation (dash-dot line) for the SPH2 simulation.
	\label{fig:massbound}
	}
	\end{center}
\end{figure}

\begin{figure}[h!]
	\begin{center}
		\includegraphics[scale=0.5]{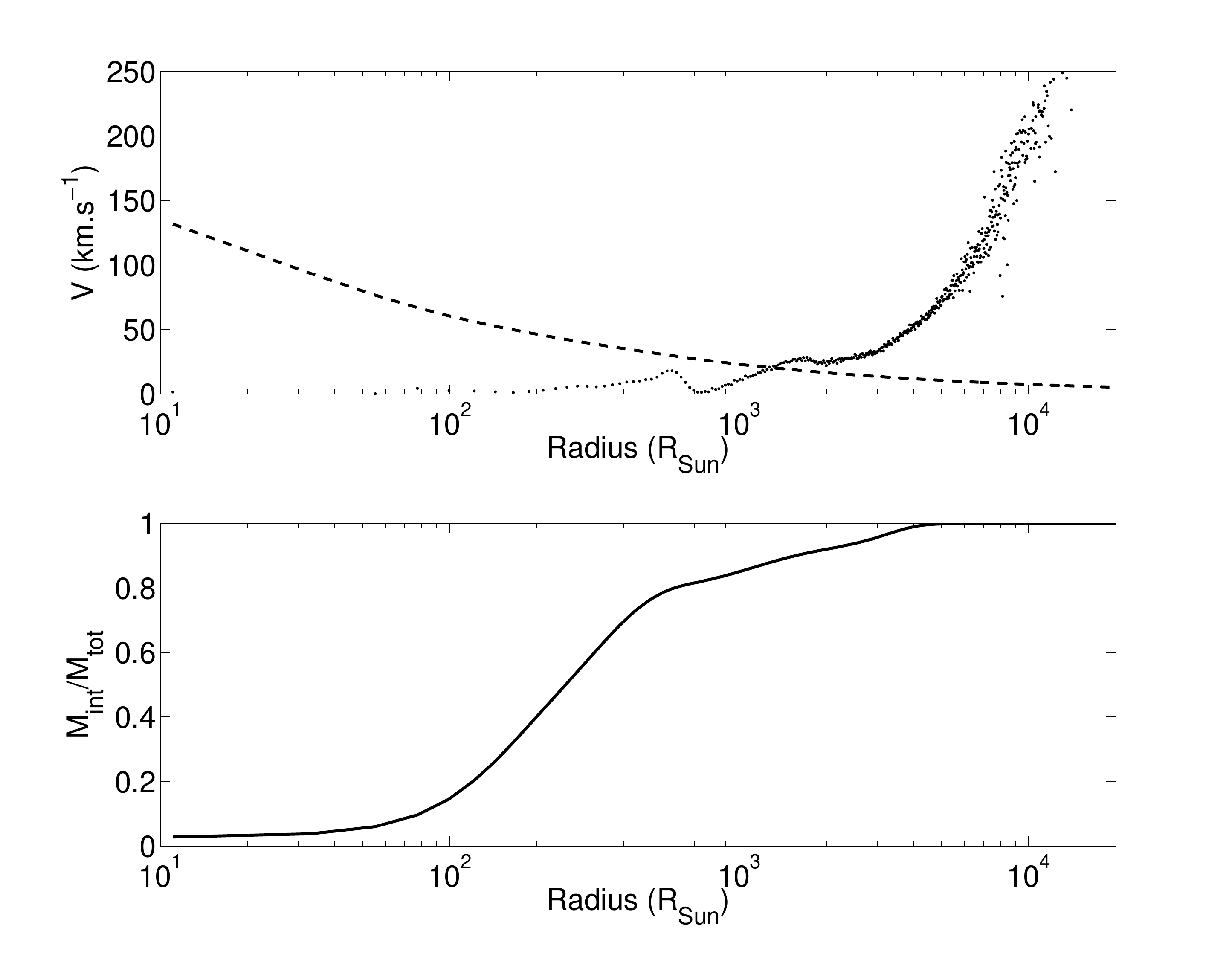}
	\caption{Top: Comparison between the escape velocity (dashed line) and the radial velocity (black dots) of the final system for the SPH2 simulation. Bottom: Mass enclosed as a function of radius.
	\label{fig:vescape}
	}
	\end{center}
\end{figure}

As shown on Table~\ref{tab:masses}, most of the primary's envelope remains bound in all of our simulations. We study the situation in detail for our canonical model with the 0.6~\msun\ companion here. The evolution of the mass for different components is plotted in Fig.~\ref{fig:massbound}. It first confirms that some envelope mass is unbound only during the first 50~days, after which neither angular momentum (Fig.~\ref{fig:AM}) nor kinetic energy (Fig.~\ref{fig:energy}) are exchanged between the unbound mass and the rest of the system. It also shows that more than 85 \% of the mass remains bound at the end of the simulation. This outcome, already pointed out by \cite{Sandquist1998}, is quite intriguing, since the post-CE binaries observed must have succeeded in ejecting their envelope. After about 400 days, most of the envelope mass in our models has been moved to a larger radius ($\sim 100$~\rsun, see bottom panel in Fig.~\ref{fig:vescape}), well outside the orbit of the primary core and the companion but remaining bound.

\begin{table}[h!]
\begin{center}
\scalebox{0.9}
	{\begin{tabular}{ccc}
	\hline
	\hline
	  & $M_2$ (\msun) & $M_{\rm bound}$ (\msun)\tablenotemark{a}\\
	\hline
	SPH1 & 0.9 & 0.44 \\
	SPH2 & 0.6 & 0.44 \\
	SPH3 & 0.3 & 0.45 \\
	SPH4 & 0.15 & 0.46 \\
	SPH5 & 0.1 & 0.48 \\
	\hline
	\hline	
 	\end{tabular}}	
	\caption{Amount of the envelope mass still bound at the end of the {\it SNSPH} simulations.
	\tablenotetext{a}{At the start of the simulations, $M_{\rm bound}$ equals the total envelope mass $M_e \equiv M_1 - M_c = 0.49$~\msun}
	\label{tab:masses}
	}
\end{center}
\end{table}

We now investigate how bound the final system is. We consider the center of mass of the system composed by the secondary and the mass within the current orbit as the center of our frame of reference. Then, we partition the domain into concentric shells with identical thickness, calculate the average radial velocity of each shell and compare it to the escape velocity at that location. Fig.~\ref{fig:vescape} shows the escape velocity and the average radial velocity of the shells. The radial velocity is always positive and is similar to the space velocity at radii larger than 600 \rsun, as expected for envelope ejection. At radii smaller than 300 \rsun, the radial velocity is much smaller than the space velocity, suggesting that orbital motions dominate at those radii. All the mass within $10^3$ \rsun \ is bound, which corresponds to more than 85\% of the envelope mass. The remaining mass is found at radii between $10^3$ and $6 \times 10^3$\rsun, where the radial velocities are typically between 25 and 75 \kms. Those particles were initially in the outer parts of the giant star, and were the first to encounter the secondary. At that time of the in-spiral, the shock was slightly supersonic ($V_2 \sim 35$~\kms \ and $c_s \sim 20$~\kms). This regime of evolution is thus different from later phases when the secondary sinks deeper into the primary's envelope, where its velocity does not really increase (Fig.~\ref{fig:velocity}) but the sound speed of the medium does (Fig.~\ref{fig:profiles}).

We can measure how much extra energy would be required to unbind the envelope at each radius, using the definition

\begin{equation}
	E_{\rm extra} = \sum_i \frac{1}{2}  M_i  \left(v_{e,i} - v_{r,i}  \right)^2
\end{equation}

\noindent where $v_{e,i}$ and $v_{r,i}$ are the escape velocity at the location of the $i$-th shell and its average radial velocity, respectively. One finds $E_{\rm extra} \sim 8.4 \times 10^{45}$ ergs which represents just over 10\% of the initial binding energy of the primary envelope. Thus, a relatively small additional input of energy could be sufficient to completely unbind the remaining envelope material.

We have compared here the final separations deduced from observations and those determined from the simulations. We have purposefully stayed away from calculating the ejection efficiency \ace \ \citep{Webbink1984, AlphaPaper2011}. Indeed, we question what the relevance of calculating \ace \ is when the envelope has not yet been fully ejected, true both in the \cite{Sandquist1998} and our simulations.  We therefore defer for the moment the task of calculating \ace \ from simulation --- a long term goal of this project --- until the simulations are more advanced.

In conclusion, the hydrodynamic simulations do not reproduce the post-CE systems in the sense that the system is left at too large separations and the envelope is not unbound at the end of the rapid infall phase. This means that either physical processes that are not accounted for in the simulations are responsible for the envelope ejection, or the envelope ejection and a significant reduction of the orbit actually happens during the later subsequent slow in-spiral phase. We discuss both possibilities in the following section.

\subsection{Reproducing the observations}
\label{subsec:reasons}

In this section we first study and quantify physical processes that are not taken into account in our hydrodynamic simulations and that might be responsible for ejecting the envelope. Then, we focus on the subsequent slow in-spiral phase and investigate whether the envelope can be ejected and the separation significantly reduced during this subsequent phase.

\subsubsection{Rotation of the primary}
\label{subsub:rotation}

The envelope of the progenitor is initially non-rotating and although the calculation done in \S\ref{sec:simus} shows that, regardless of the initial rotation velocity of the envelope, its rotational energy is negligible in comparison with its binding energy, we suspected at first that the absence of rotation might be the reason for most of the envelope to remain bound. However, \cite{Sandquist1998} carried out two identical simulations where they modified the initial rotation state of the primary from a giant star in synchronization with the orbit to a non-rotating one (their simulations 1 and 2). In both cases, the evolution of the bound mass and the final orbital parameters are similar.  It thus does not seem that changing the initial rotation of the primary leads to a different CE outcome. 

\subsubsection{Physics not included in the simulations}
\label{subsub:missing}

The hydrodynamics codes use an ideal gas equation of state (\S\ref{subsec:codesdiff}) which, by definition, does not include variable abundances and the different ionization layers of the envelope. \cite{Han1995} suggested that recombination might play a role in CE interactions. As the outer parts of the envelope expand and cool, ions recombine with electrons, releasing energy that could aid in unbinding the envelope. Although it is unclear how efficient this process is and how much of the initial recombination budget can be used, one can calculate an upper limit on how much energy can be injected into the envelope by recombination. 

According to our stellar evolution model, the hydrogen fraction within the convective envelope of our RGB star is $X \sim 0.68$. The mass of the envelope is $M_e = 0.49$~\msun \ and each proton recombining with an electron produces an energy $E_0 = 13.6$~eV. We also have to calculate how much of the envelope is ionized. Therefore, we calculate the partition functions $Z$ for hydrogen. The hydrogen ion has no degeneracy so $Z_2 = 1$. The partition function for the hydrogen atom at temperature $T$ is

\begin{equation}
	Z_1 = \sum_{n=1}^{\infty} 2n^2 \exp{\frac{E_0(1/n^2-1)}{kT}}
	\label{eq:partition}
\end{equation}

\noindent where $k=8.6173 \times 10^{-5}$~eV.$K^{-1}$ is the Boltzmann constant. We truncate the sum in Eq.~\ref{eq:partition} at the first integer $n_{\rm max}$ such that the distance at which the electron orbits the proton for this quantum number is larger than $l_{\rm max}=10^{-6}$~cm, i.e. $a_0n_{\rm max}^2 > l_{\rm max}$ where $a_0 = 5.2918 \times 10^9$~cm is the Bohr radius \citep{Miranda2001}. We then use the Saha formula to calculate the ratio of ionized to neutral hydrogen \citep{Carroll}:

\begin{equation}
	N_2/N_1 = \frac{2Z_2}{n_e Z_1}\left(\frac{2\pi m_e k T}{h^2}\right)^{3/2} \exp{(-E_0/kT)}
\end{equation}

\noindent where $n_e$ is the number density of free electrons and $m_e$ is the electron mass. We find that 91\% of the envelope is ionized. Consequently, the recombination of the whole ionized envelope would produce an extra energy

\begin{equation}
	E_{\rm recomb} = 0.91 \times XM_e \frac{N_A}{M_H} \times 13.6~{\rm eV}
	\label{eq:recomb}
\end{equation}

\noindent where $N_A$ is the Avogadro number and $M_H$ the atomic mass of hydrogen. One finds $E_{\rm recomb} \sim 1.18 \times 10^{46}$~ergs, which is slightly higher than the extra energy $E_{\rm extra}$ required to eject the envelope in our canonical model (\S\ref{sec:envelope}). Thus, we conclude that recombination in the envelope could substantially aid in unbinding it.

Another source of energy could be radiation pressure. For low- and intermediate-mass giants in hydrostatic equilibrium, radiation pressure ($P_{rad} \equiv a T^4/3$, where $a$ is the radiation constant) is negligible compared to gas pressure (Eq.~\ref{eq:hd4}): for our primary, $P_{rad} / P_{gas} \la 0.01$ except in a small zone ($0.1$ \rsun $ \leq r \leq 10$ \rsun), where $P_{rad} / P_{gas} \la 0.1$. However, the deep in-spiral of the companion within the primary's envelope will induce local shock heating. The increase of temperature is proportional to the square of the Mach number \citep{TarbellEtAl1999}, so even if the companion is orbiting at twice the local sound speed, the radiation pressure to  gas pressure after the shock becomes:

\begin{equation}
	\left( \frac{P_{rad}}{P_{gas}} \right)^{\rm after} \propto \left( \frac{P_{rad}}{P_{gas}} \right)^{\rm before} (M^2)^3 = 6.4
\end{equation}

\noindent Therefore, including radiation pressure in the equation of state will increase the total pressure locally and might reduce the energy required to eject the envelope. However, it is possible that this effect is globally small, since this extra heating source is probably very localized around the companion. 

\subsubsection{The post-rapid-infall phase}

At the end of the rapid infall phase, the orbit is stable until the end of the simulations (a few more years). Consequently, there is no further hydrodynamical coupling between the extended envelope and the surviving binary. We now investigate whether the envelope is likely to be ejected during this slower in-spiral phase.

Although the resolution of the simulations prevents us from quantifying how much envelope will be left around the core of the primary, one can still describe qualitatively what the evolution of the primary's remnant will be. Fig.~\ref{fig:massbound} shows that less than $10^{-2}$~\msun \ is left around the primary's core, so the primary will depart the giant branch (\citealt{Bloecker1995}, but see also the discussion in \citealt{AlphaPaper2011}). Then two scenarios might occur depending on how long the partially ejected envelope will take to fall back.

If the star is given enough time to transit to the blue due to hydrogen burning at the base of the envelope before the lifted envelope falls back, the star will readjust on its thermal timescale of the remaining envelope, and eventually end its life as a Helium white dwarf. This transition will last $\sim 10^3$~years during which the star will have a luminosity between 300 and 1\,000 \lsun \ \citep[][their Fig.~1]{IbenTutukov1993}, which is consistent with the more recent work of \cite{DriebeEtAl1998} (their Fig.~1). If we assume the remnant to have a luminosity $L_c \sim 500$~\lsun, we can compare the gravitational acceleration of a gas particle with the radiation acceleration defined by

\begin{equation}
	a_{rad} = \frac{L_c}{4\pi r^2} \frac{\kappa}{c} 
	\label{eq:radacc}
\end{equation}

\noindent where $r$ is the distance between the gas particle and the core, $\kappa = 0.4$~cm$^2$\,$g^{-1}$ is the opacity for Thompson scattering for hydrogen, and $c$ is the speed of light. We still find the radiation acceleration to be overall almost two orders of magnitude smaller than the gravitational acceleration. 

If on the contrary, the envelope falls back before the primary's remnant had crossed the Hertzsprung-Russell diagram, a circumbinary disk will form \citep{KashiSoker2011}. They refer to the numerical work done by \cite{ArtymowiczEtAl1991}, which suggests that in such a configuration, the binary separation will decrease due to Lindblad resonances --- mainly --- as well as viscous tides. Although this mechanism has the advantage of explaining how the orbital separation will diminish during the subsequent phase, the ability of radiation to eject the gas will even be reduced in comparison with the previous situation, so it is not clear how the latter will eventually be unbound.

In conclusion, radiation acceleration alone does not seem to be responsible for unbinding the remaining gas, regardless of the time the partially ejected envelope will remain suspended for.

\section{Summary and Future Work}
\label{sec:conclu}

In this work we have carried out three-dimensional hydrodynamic simulations of the CE interaction between a 0.88~\msun \ RGB star and companions with mass ranging from 0.1 to 0.9~\msun. We have used both an Eulerian grid code ({\it Enzo}) and a Lagrangian SPH code ({\it SNSPH}) with various resolutions. They both have advantages and disadvantages and can be used for different purposes: while one might rather use SPH to study the accretion around the secondary, even a uniform grid code is more suitable in resolving the low-density extended envelope. Of course, adaptive mesh refinement combines the advantages of both of these methods at the cost of increased code complexity.

We first compared the outcomes of those simulations with each other. We found that the results are very similar for companion masses $M_2 \ga 0.3$~\msun. We thus conclude that in this regime, the resolutions used are sufficient to study the global evolution of the system during the rapid infall phase of the interaction, which is driven mainly by gravitational drag. For lower companion masses ($M_2 \la 0.3$~\msun) that penetrate deeper in the giant's envelope, the $128^3$ {\it Enzo} runs are under-resolved but the {\it Enzo} results converge to the solutions from the  {\it SNSPH} simulations.

We then compared the outcomes of our simulations with observed post-CE systems. The final separations are found to be systematically higher than those deduced from observations, as is the case for the past simulations by \cite{Sandquist1998}, \cite{DeMarco2003} and \cite{RickerTaam2008}. Moreover, mass is only unbound during the early stages of the interaction ($\sim 50$~days for the 0.6 \msun \ companion) and most of the envelope remains bound at the end of the simulations, as was the case for the earlier simulations of \cite{Sandquist1998}. We investigated whether there might be additional processes that were not accounted for in the simulations. We found that recombination can contribute significantly, but stellar rotation and radiation pressure play only marginal roles. Finally, we wondered whether the bound envelope is a result of  imprecise simulations or a real physical feature. If the latter, then one would have to follow the subsequent evolution of the system to determine the actual outcome of the CE. Fall back disks may form and even have an impact on the inner binary \citep[][Kashi \& Soker 2011]{ArtymowiczEtAl1991}.

After the submission of this paper, Ricker and Taam made their paper \cite{RickerTaam2011} available. This paper continues the work introduced in \cite{RickerTaam2008}. In their simulation, only about 25 \% of the primary's envelope is unbound. Although this value is slightly higher than ours, it is in agreement with our work in the sense that most of the envelope remains bound. They also claim that the ejection occurs mostly in the orbital plane, as it is the case in our simulations. However, the extended envelope at the end of their simulation is rotating much faster than it is expanding which is in contradiction with our results (\S~\ref{sec:envelope}) but might be due to the fact that their primary is initially rotating.
 
\section{Acknowledgments}
\label{sec:ack}
J-CP, ODM and M-MML acknowledge funding from NSF grant 0607111. FH acknowledges funding from an NSERC Discovery grant. J-CP thanks Colin McNally for useful comments and discussions. J-CP is grateful to both the {\it Enzo} and {\it yt} \citep{YtPaper} communities for their precious help, especially Matthew Turk's dedication. The authors acknowledge computer time provided by Westgrid and Compute Canada as well as the Los Alamos National Laboratory. The authors thank an anonymous referee for his useful comments that helped to improve this manuscript.

\newpage
\bibliographystyle{/Applications/TeX/apj}                       
\bibliography{/Applications/TeX/bibliography}

\begin{thebibliography}{57}
\expandafter\ifx\csname natexlab\endcsname\relax\def\natexlab#1{#1}\fi

\bibitem[{{Agertz} {et~al.}(2007){Agertz}, {Moore}, {Stadel}, {Potter},
  {Miniati}, {Read}, {Mayer}, {Gawryszczak}, {Kravtsov}, {Nordlund}, {Pearce},
  {Quilis}, {Rudd}, {Springel}, {Stone}, {Tasker}, {Teyssier}, {Wadsley}, \&
  {Walder}}]{AgertzEtAl2007}
{Agertz}, O., {Moore}, B., {Stadel}, J., {Potter}, D., {Miniati}, F., {Read},
  J., {Mayer}, L., {Gawryszczak}, A., {Kravtsov}, A., {Nordlund}, {\AA}.,
  {Pearce}, F., {Quilis}, V., {Rudd}, D., {Springel}, V., {Stone}, J.,
  {Tasker}, E., {Teyssier}, R., {Wadsley}, J., \& {Walder}, R. 2007, \mnras,
  380, 963

\bibitem[{{Artymowicz} {et~al.}(1991){Artymowicz}, {Clarke}, {Lubow}, \&
  {Pringle}}]{ArtymowiczEtAl1991}
{Artymowicz}, P., {Clarke}, C.~J., {Lubow}, S.~H., \& {Pringle}, J.~E. 1991,
  \apjl, 370, L35

\bibitem[{{Belczynski} {et~al.}(2008){Belczynski}, {Kalogera}, {Rasio}, {Taam},
  {Zezas}, {Bulik}, {Maccarone}, \& {Ivanova}}]{Belczynski2008}
{Belczynski}, K., {Kalogera}, V., {Rasio}, F.~A., {Taam}, R.~E., {Zezas}, A.,
  {Bulik}, T., {Maccarone}, T.~J., \& {Ivanova}, N. 2008, \apjs, 174, 223

\bibitem[{{Benz}(1989)}]{Benz1989}
{Benz}, W. 1989, in The Numerical Modeling of Stellar Pulsation, ed. J. R.
  Buchler (Dordrecht: Kluwer), 269

\bibitem[{{Bloecker}(1995)}]{Bloecker1995}
{Bloecker}, T. 1995, \aap, 299, 755

\bibitem[{{Bodenheimer} \& {Taam}(1984)}]{PaperII}
{Bodenheimer}, P. \& {Taam}, R.~E. 1984, \apj, 280, 771

\bibitem[{{Bondi} \& {Hoyle}(1944)}]{BondiHoyle1944}
{Bondi}, H. \& {Hoyle}, F. 1944, \mnras, 104, 273

\bibitem[{{Bryan} {et~al.}(1995){Bryan}, {Norman}, {Stone}, {Cen}, \&
  {Ostriker}}]{BryanEtAl1995}
{Bryan}, G.~L., {Norman}, M.~L., {Stone}, J.~M., {Cen}, R., \& {Ostriker},
  J.~P. 1995, Computer Physics Communications, 89, 149

\bibitem[{{Carroll} \& {Ostlie}(1996)}]{Carroll}
{Carroll}, B.~W. \& {Ostlie}, D.~A. 1996, {An Introduction to Modern
  Astrophysics} (Institute for Mathematics and Its Applications)

\bibitem[{{Darwin}(1879)}]{Darwin1879}
{Darwin}, G.~H. 1879, The Observatory, 3, 79

\bibitem[{{Davies} {et~al.}(1993){Davies}, {Ruffert}, {Benz}, \&
  {Muller}}]{DaviesEtAl1993}
{Davies}, M.~B., {Ruffert}, M., {Benz}, W., \& {Muller}, E. 1993, \aap, 272,
  430

\bibitem[{{De Marco} {et~al.}(2011){De Marco}, {Passy}, {Moe}, {Herwig}, {Mac
  Low}, \& {Paxton}}]{AlphaPaper2011}
{De Marco}, O., {Passy}, {\relax J.-C}., {Moe}, M., {Herwig}, F., {Mac Low},
  {\relax M.-M}., \& {Paxton}, B. 2011, \mnras, 28

\bibitem[{{De Marco} {et~al.}(2003){De Marco}, {Sandquist}, {Mac Low},
  {Herwig}, \& {Taam}}]{DeMarco2003}
{De Marco}, O., {Sandquist}, E.~L., {Mac Low}, M.-M., {Herwig}, F., \& {Taam},
  R.~E. 2003, in Rev. Mex. Astron. Astrof. Conf. Ser., 24--30

\bibitem[{{de Medeiros} \& {Mayor}(1999)}]{DeMedeirosMayor1999}
{de Medeiros}, J.~R. \& {Mayor}, M. 1999, \aaps, 139, 433

\bibitem[{{Diehl} \& {Statler}(2006)}]{DiehlStatler2006}
{Diehl}, S. \& {Statler}, T.~S. 2006, \mnras, 368, 497

\bibitem[{{Driebe} {et~al.}(1998){Driebe}, {Schoenberner}, {Bloecker}, \&
  {Herwig}}]{DriebeEtAl1998}
{Driebe}, T., {Schoenberner}, D., {Bloecker}, T., \& {Herwig}, F. 1998, \aap,
  339, 123

\bibitem[{{Duquennoy} \& {Mayor}(1991)}]{Duquennoy1991}
{Duquennoy}, A. \& {Mayor}, M. 1991, \aap, 248, 485

\bibitem[{{Frenk} {et~al.}(1999){Frenk}, {White}, {Bode}, {Bond}, {Bryan},
  {Cen}, {Couchman}, {Evrard}, {Gnedin}, {Jenkins}, {Khokhlov}, {Klypin},
  {Navarro}, {Norman}, {Ostriker}, {Owen}, {Pearce}, {Pen}, {Steinmetz},
  {Thomas}, {Villumsen}, {Wadsley}, {Warren}, {Xu}, \& {Yepes}}]{SantaBarbara}
{Frenk}, C.~S., {White}, S.~D.~M., {Bode}, P., {Bond}, J.~R., {Bryan}, G.~L.,
  {Cen}, R., {Couchman}, H.~M.~P., {Evrard}, A.~E., {Gnedin}, N., {Jenkins},
  A., {Khokhlov}, A.~M., {Klypin}, A., {Navarro}, J.~F., {Norman}, M.~L.,
  {Ostriker}, J.~P., {Owen}, J.~M., {Pearce}, F.~R., {Pen}, U., {Steinmetz},
  M., {Thomas}, P.~A., {Villumsen}, J.~V., {Wadsley}, J.~W., {Warren}, M.~S.,
  {Xu}, G., \& {Yepes}, G. 1999, \apj, 525, 554

\bibitem[{{Fryer} {et~al.}(2006){Fryer}, {Rockefeller}, \&
  {Warren}}]{FryerEtAl2006}
{Fryer}, C.~L., {Rockefeller}, G., \& {Warren}, M.~S. 2006, \apj, 643, 292

\bibitem[{{Fryxell} {et~al.}(2000){Fryxell}, {Olson}, {Ricker}, {Timmes},
  {Zingale}, {Lamb}, {MacNeice}, {Rosner}, {Truran}, \& {Tufo}}]{FlashCode}
{Fryxell}, B., {Olson}, K., {Ricker}, P., {Timmes}, F.~X., {Zingale}, M.,
  {Lamb}, D.~Q., {MacNeice}, P., {Rosner}, R., {Truran}, J.~W., \& {Tufo}, H.
  2000, \apjs, 131, 273

\bibitem[{{Han} {et~al.}(1995){Han}, {Podsiadlowski}, \& {Eggleton}}]{Han1995}
{Han}, Z., {Podsiadlowski}, P., \& {Eggleton}, P.~P. 1995, \mnras, 272, 800

\bibitem[{{Heitsch} {et~al.}(2011){Heitsch}, {Naab}, \&
  {Walch}}]{HeitschEtAl2011}
{Heitsch}, F., {Naab}, T., \& {Walch}, S. 2011, ArXiv e-prints

\bibitem[{{Herwig}(2000)}]{Herwig2000}
{Herwig}, F. 2000, \aap, 360, 952

\bibitem[{{Hjellming} \& {Webbink}(1987)}]{HjellmingWebbink1987}
{Hjellming}, M.~S. \& {Webbink}, R.~F. 1987, \apj, 318, 794

\bibitem[{{Hurley} {et~al.}(2002){Hurley}, {Tout}, \& {Pols}}]{HurleyEtAl2002}
{Hurley}, J.~R., {Tout}, C.~A., \& {Pols}, O.~R. 2002, \mnras, 329, 897

\bibitem[{{Iben} \& {Tutukov}(1993)}]{IbenTutukov1993}
{Iben}, Jr., I. \& {Tutukov}, A.~V. 1993, \apj, 418, 343

\bibitem[{{Iben} \& {Livio}(1993)}]{Iben1993}
{Iben}, I.~J. \& {Livio}, M. 1993, \pasp, 105, 1373

\bibitem[{{Kashi} \& {Soker}(2011)}]{KashiSoker2011}
{Kashi}, A. \& {Soker}, N. 2011, ArXiv e-prints

\bibitem[{{Maxted} {et~al.}(2006){Maxted}, {Napiwotzki}, {Dobbie}, \&
  {Burleigh}}]{Maxted2006}
{Maxted}, P.~F.~L., {Napiwotzki}, R., {Dobbie}, P.~D., \& {Burleigh}, M.~R.
  2006, \nat, 442, 543

\bibitem[{{Meng} {et~al.}(2010){Meng}, {Chen}, {Yang}, \& {Li}}]{MengEtAl2010}
{Meng}, X., {Chen}, W., {Yang}, W., \& {Li}, Z. 2010, ArXiv e-prints

\bibitem[{{Miranda}(2001)}]{Miranda2001}
{Miranda}, E.~N. 2001, European Journal of Physics, 22, 483

\bibitem[{{Monaghan}(1992)}]{Monaghan1992}
{Monaghan}, J.~J. 1992, \araa, 30, 543

\bibitem[{{Nelemans} {et~al.}(2000){Nelemans}, {Verbunt}, {Yungelson}, \&
  {Portegies Zwart}}]{Nelemans2000}
{Nelemans}, G., {Verbunt}, F., {Yungelson}, L.~R., \& {Portegies Zwart}, S.~F.
  2000, \aap, 360, 1011

\bibitem[{{Norman} {et~al.}(2007){Norman}, {Bryan}, {Harkness}, {Bordner},
  {Reynolds}, {O'Shea}, \& {Wagner}}]{NormanEtAl2007}
{Norman}, M.~L., {Bryan}, G.~L., {Harkness}, R., {Bordner}, J., {Reynolds}, D.,
  {O'Shea}, B., \& {Wagner}, R. 2007, ArXiv e-prints

\bibitem[{{O'Shea} {et~al.}(2004){O'Shea}, {Bryan}, {Bordner}, {Norman},
  {Abel}, {Harkness}, \& {Kritsuk}}]{OsheaEtAl2004}
{O'Shea}, B.~W., {Bryan}, G., {Bordner}, J., {Norman}, M.~L., {Abel}, T.,
  {Harkness}, R., \& {Kritsuk}, A. 2004, ArXiv Astrophysics e-prints

\bibitem[{{Paczynski}(1976)}]{Paczynski1976}
{Paczynski}, B. 1976, in IAU Symp. 73: Structure and Evolution of Close Binary
  Systems, ed. P.~{Eggleton}, S.~{Mitton}, \& J.~{Whelan}, 75--+

\bibitem[{{Perets}(2010)}]{Perets2010}
{Perets}, H.~B. 2010, ArXiv e-prints

\bibitem[{{Podsiadlowski}(2001)}]{Podsiadlowski2001}
{Podsiadlowski}, P. 2001, in Astronomical Society of the Pacific Conference
  Series, Vol. 229, Evolution of Binary and Multiple Star Systems, ed.
  {P.~Podsiadlowski, S.~Rappaport, A.~R.~King, F.~D'Antona, \& L.~Burderi },
  239--+

\bibitem[{{Politano} {et~al.}(2010){Politano}, {van der Sluys}, {Taam}, \&
  {Willems}}]{PolitanoEtAl2010}
{Politano}, M., {van der Sluys}, M., {Taam}, R.~E., \& {Willems}, B. 2010,
  \apj, 720, 1752

\bibitem[{{Ricker} \& {Taam}(2008)}]{RickerTaam2008}
{Ricker}, P.~M. \& {Taam}, R.~E. 2008, \apjl, 672, L41

\bibitem[{{Ricker} \& {Taam}(2011)}]{RickerTaam2011}
---. 2011, ArXiv e-prints

\bibitem[{{Rosswog}(2009)}]{Rosswog2009}
{Rosswog}, S. 2009, \nar, 53, 78

\bibitem[{{Ruffert}(1993)}]{Ruffert1993}
{Ruffert}, M. 1993, \aap, 280, 141

\bibitem[{{Sandquist} {et~al.}(2000){Sandquist}, {Taam}, \&
  {Burkert}}]{SandquistEtAl2000}
{Sandquist}, E.~L., {Taam}, R.~E., \& {Burkert}, A. 2000, \apj, 533, 984

\bibitem[{{Sandquist} {et~al.}(1998){Sandquist}, {Taam}, {Chen}, {Bodenheimer},
  \& {Burkert}}]{Sandquist1998}
{Sandquist}, E.~L., {Taam}, R.~E., {Chen}, X., {Bodenheimer}, P., \& {Burkert},
  A. 1998, \apj, 500, 909

\bibitem[{{Schreiber} \& {G{\"a}nsicke}(2003)}]{Schreiber2003}
{Schreiber}, M.~R. \& {G{\"a}nsicke}, B.~T. 2003, \aap, 406, 305

\bibitem[{{Setiawan} {et~al.}(2010){Setiawan}, {Klement}, {Henning}, {Rix},
  {Rochau}, {Rodmann}, \& {Schulze-Hartung}}]{SetiawanEtAl2010}
{Setiawan}, J., {Klement}, R.~J., {Henning}, T., {Rix}, H.-W., {Rochau}, B.,
  {Rodmann}, J., \& {Schulze-Hartung}, T. 2010, Science, 330, 1642

\bibitem[{{Stone} \& {Norman}(1992)}]{StoneNorman1992}
{Stone}, J.~M. \& {Norman}, M.~L. 1992, \apjs, 80, 753

\bibitem[{{Taam} \& {Ricker}(2010)}]{TaamRicker2010}
{Taam}, R.~E. \& {Ricker}, P.~M. 2010, \nar, 54, 65

\bibitem[{{Taam} \& {Sandquist}(2000)}]{TaamSandquist2000}
{Taam}, R.~E. \& {Sandquist}, E.~L. 2000, \araa, 38, 113

\bibitem[{{Tarbell} {et~al.}(1999){Tarbell}, {Ryutova}, {Covington}, \&
  {Fludra}}]{TarbellEtAl1999}
{Tarbell}, T., {Ryutova}, M., {Covington}, J., \& {Fludra}, A. 1999, \apjl,
  514, L47

\bibitem[{{Tasker} {et~al.}(2008){Tasker}, {Brunino}, {Mitchell}, {Michielsen},
  {Hopton}, {Pearce}, {Bryan}, \& {Theuns}}]{TaskerEtAl2008}
{Tasker}, E.~J., {Brunino}, R., {Mitchell}, N.~L., {Michielsen}, D., {Hopton},
  S., {Pearce}, F.~R., {Bryan}, G.~L., \& {Theuns}, T. 2008, \mnras, 390, 1267

\bibitem[{{Turk} {et~al.}(2011){Turk}, {Smith}, {Oishi}, {Skory}, {Skillman},
  {Abel}, \& {Norman}}]{YtPaper}
{Turk}, M.~J., {Smith}, B.~D., {Oishi}, J.~S., {Skory}, S., {Skillman}, S.~W.,
  {Abel}, T., \& {Norman}, M.~L. 2011, \apjs, 192, 9

\bibitem[{{van Leer}(1977)}]{VanLeer1977}
{van Leer}, B. 1977, J. Comput. Phys., 23, 276

\bibitem[{Warren \& Salmon(1993)}]{WarrenSalmon1993}
Warren, M.~S. \& Salmon, J.~K. 1993, in Supercomputing '93: Proceedings of the
  1993 ACM/IEEE conference on Supercomputing (New York, NY, USA: ACM), 12--21

\bibitem[{{Webbink}(1984)}]{Webbink1984}
{Webbink}, R.~F. 1984, \apj, 277, 355

\bibitem[{{Zorotovic} {et~al.}(2010){Zorotovic}, {Schreiber}, {G{\"a}nsicke},
  \& {Nebot G{\'o}mez-Mor{\'a}n}}]{Zorotovic2010}
{Zorotovic}, M., {Schreiber}, M.~R., {G{\"a}nsicke}, B.~T., \& {Nebot
  G{\'o}mez-Mor{\'a}n}, A. 2010, \aap, 520, A86+

\end{thebibliography}

\end{document}